\documentclass[12pt]{article}

\usepackage[textheight=22.5truecm,textwidth=16.8truecm,top=2.95truecm,left=2.15truecm]{geometry}
\usepackage{
amsmath,
amsthm,
amssymb,
ascmac,
bm,
tikz,
braket,
mathrsfs,
graphicx,
color,
hyperref,
cite,
titlesec,
authblk,
caption,
enumitem,
cancel,
here
}

\hypersetup{colorlinks,bookmarksopen,bookmarksnumbered,citecolor=black,linkcolor=black,linktocpage,pdfstartview=FitH,urlcolor=black}

\numberwithin{equation}{section}

{\makeatletter
\long\def\@makefntext#1{\parindent 1em\noindent 
\@hangfrom{\hbox to 1.8em{\hss$^{\@thefnmark}$}}#1}
\makeatother}
\renewcommand\thefootnote{\arabic{footnote})}
\usepackage[hang,flushmargin,bottom]{footmisc}

\def\fnum@figure{\textbf{\figurename\nobreakspace\thefigure}}
\def\fnum@table{\textbf{\tablename\nobreakspace\thetable}}
\long\def\@makecaption#1#2{%
  \vskip\abovecaptionskip
  \sbox\@tempboxa{\small #1. #2}%
  \ifdim \wd\@tempboxa >\hsize
    \small #1. #2\par
  \else
    \global \@minipagefalse
    \hb@xt@\hsize{\hfil\box\@tempboxa\hfil}%
  \fi
  \vskip\belowcaptionskip}
\renewcommand{\l}[0]{\left}
\renewcommand{\r}[0]{\right}
\renewcommand{\d}[0]{d}

\newcommand{\cmt}[2]{[#1,#2]}

\renewcommand{\.}[0]{\hspace{0mm}}

\newcommand{\ttl}[0]{\mathrm{tot}}
\newcommand{\ith}[1]{{(#1)}}

\newcommand{\nn}{\nonumber\\}
\newcommand{\bz}{\bar{z}}

\DeclareMathOperator{\Tr}{Tr}
\DeclareMathOperator{\tr}{Tr}
\newcommand{\eqb}[1]{\braket{#1}_\beta}
\newcommand{\cO}{\mathcal{O}}

\allowdisplaybreaks

\title{\hfill\parbox{3cm}{
\normalsize
KUNS-3031
}\\[12pt]
Heat and work in black hole thermodynamics\\ via holography
}

\author{Tomohiro Shigemura, Keito Shimizu, Sotaro Sugishita,\\ Daichi Takeda, Takuya Yoda}

\affil{\it Department of Physics, Kyoto University, Kyoto 606-8502, Japan}

\date{}

\begin{document}
\maketitle

\renewcommand{\thefootnote}{\fnsymbol{footnote}}
\footnote[0]{
shigemura(at)gauge.scphys.kyoto-u.ac.jp,
kate(at)gauge.scphys.kyoto-u.ac.jp,
sotaro(at)gauge.scphys.kyoto-u.ac.jp, 
takedai(at)gauge.scphys.kyoto-u.ac.jp,
t.yoda(at)gauge.scphys.kyoto-u.ac.jp
}
\renewcommand{\thefootnote}{\arabic{footnote}}

\begin{abstract}
We propose a formulation of black hole thermodynamics that incorporates the notions of heat and work, based on the thermodynamics in quantum theory and the AdS/CFT correspondence. First, for coupled holographic CFTs, we define a coarse-graining procedure adopting the principle of maximum entropy. Employing this approach, when the system is divided into a target system and thermal baths, we formulate the first and second laws, as well as the fundamental thermodynamic relation. Then, by translating the resulting thermodynamics into the AdS gravity language, we construct a thermodynamic framework for composite black hole systems that encompasses both heat and work. This formulation relies on holography, but not on energy conditions on the gravity side. We also argue that the second law serves as a necessary criterion for the UV completeness of gravitational theories.
\end{abstract}

\newpage
\tableofcontents
\newpage

\section{Introduction}
\label{sec:intro}
Thermodynamic properties of black holes imply the existence of some unknown microscopic theory behind general relativity.
The attempts of black hole microstate counting initiated by \cite{Strominger:1996sh} have provided compelling evidence for this idea.
In the context of the AdS/CFT correspondence \cite{Maldacena:1997re}, the entropy of black holes coincides with the thermal entropy of the corresponding CFT states \cite{Gubser:1996de, Witten:1998zw}, and this fact also implies the existence of microscopic degrees of freedom responsible for the black hole entropy.
In light of these facts, the completion of black hole thermodynamics is one of the most intriguing topics to approach quantum gravity.

An issue that needs to be addressed is that the generic notions of heat and work are missing from black hole thermodynamics. 
In general, heat and work play indispensable roles in thermodynamics.
A concept of thermodynamic work was already introduced in black hole thermodynamics \cite{Johnson:2014yja} by regarding the cosmological constant as a thermodynamic parameter (see also \cite{Teitelboim:1985dp, Caldarelli:1999xj, Sekiwa:2006qj, Kastor:2009wy, Dolan:2010ha, Kubiznak:2014zwa}).
It is indeed natural to regard the cosmological constant as pressure (i.e., the conjugate variable of volume) \cite{Kastor:2009wy}.
Then, we can consider a thermodynamic cycle (e.g., the Carnot cycle) on the pressure-volume plane, and also define the heat associated with the cycle \cite{Johnson:2014yja}.
However, the notion of heat and work is more generic in thermodynamics, not restricted to the change of energy due to the change of pressure and volume.
It is desirable to have a more general notion of heat and work in black hole thermodynamics.
In addition, as also mentioned in \cite{Johnson:2014yja}, changing the cosmological constant involves changing the theory. 
Indeed, in AdS/CFT, the change corresponds to the change of the gauge rank $N$, if the dual field theory is an SU($N$) gauge theory.
Since, in the standard thermodynamics, we can introduce work and heat without changing the underlying microscopic theory, for example, by applying external sources, we should be able to do in black hole thermodynamics. 
We need to fill in this gap to complete the construction of black hole thermodynamics.
It will also provide clues to the evolution of interacting black holes.

Another essential problem toward the completion of black hole thermodynamics is extending it to non-equilibrium thermodynamics. 
Recent developments in non-equilibrium thermodynamics suggest that we can define a notion of entropy even for non-equilibrium states.  
Thus, it is natural to expect that we can define entropy for dynamical black holes and to ask whether it follows the second law.
There are attempts to define entropy for dynamical black holes (see, e.g., \cite{Iyer:1994ys, Dong:2013qoa, Wall:2015raa, Hollands:2024vbe}). 
In general, it is believed that we have the generalized second law in gravitational systems \cite{Bekenstein:1973ur}: the generalized entropy, the sum of the black hole entropy and the entropy of matter outside the black hole, never decreases. 
While the black hole entropy is thought to be given by a surface area (for the Einstein gravity), the appropriate choice of the surface, such as the event horizon or the apparent horizon, is still under debate.
Also, the choice of energy condition to prove the second law is still ambiguous.
For references of the debates, see \cite{Wall:2009wm, Carlip:2014pma, Wall:2018ydq, Sarkar:2019xfd}.

However, these problems with the second law are inherently difficult to solve without any access to the microscopic definition of entropy.
For example, in stochastic thermodynamics, we have a time-evolving probability distribution $p_x$ ($x$ is the label of the states), and its entropy is defined as the Shannon entropy, $-\sum_x p_x \ln p_x$.
From these, we can derive the second law or the fluctuation theorem \cite{evans1993probability,kurchan1998fluctuation}.
Similarly, in quantum thermodynamics,
we have the von Neumann equation or the master equation for a state $\rho(t)$, and the entropy can be defined as the von Neumann entropy, $-\Tr\rho(t) \ln \rho(t)$.
On the other hand, we do not know the true microscopic degrees of freedom of gravity yet.

Our approach to overcome the difficulties is to use the AdS/CFT correspondence, which in principle, provides a microscopic definition of quantum gravity. 
It will lead to a proper definition of entropy even for dynamical black holes and the way to follow its time evolution, since the dual field theory is expected to know the bulk quantum gravity.
Among attempts \cite{Kelly:2013aja,Kelly:2014owa,Hubeny:2012wa,Freivogel:2013zta, Engelhardt:2017wgc,Engelhardt:2017aux,Grado-White:2017nhs,Engelhardt:2018kcs,Chandra:2022fwi},
one of the authors proposed a new thermodynamic framework in \cite{Takeda:2024qbq}.
In the work, the bulk definition of entropy is automatically determined from the coarse-grained entropy introduced by considering the thermodynamics of the boundary theory.
The second law is first derived on the boundary side, and then, it is translated into the bulk language using the AdS/CFT dictionary \cite{Gubser:1998bc, Witten:1998qj}.
The entropy is free from an artificial choice of spacetime surfaces and is independent of bulk time foliations.
Based on the holographic principle, the bulk second law is shown from the non-negativity of the relative entropy\footnote{The relative entropy is often used in quantum thermodynamics (see, e.g., \cite{Vedral:2002zz, Sagawa:2012eqh} for reviews). 
It is also used to derive various properties of entropy and energy conditions in systems with gravity \cite{Casini:2008cr, Wall:2011hj, Bousso:2014sda, Faulkner:2016mzt}, and applied to AdS/CFT \cite{Blanco:2013joa, Lashkari:2013koa, Faulkner:2013ica, Banerjee:2014oaa, Banerjee:2014ozp, Lashkari:2014kda, Jafferis:2015del} as a tool to survey the bulk from the boundary point of view.}
among the CFT states --- there is no room for choosing energy conditions.

\begin{figure}[t]
	\centering
	\includegraphics[width = 0.6\columnwidth]{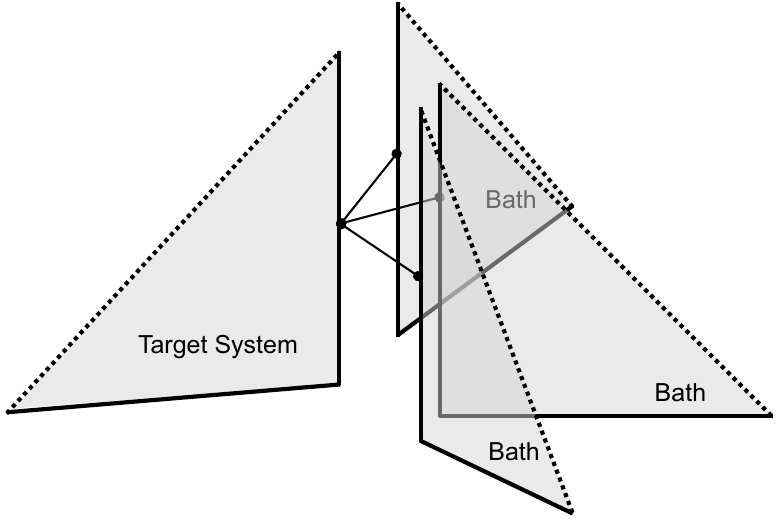}
	\caption{The system and baths consisting of single-sided asymptotically AdS black holes, for which we construct thermodynamics.
	Only the segment $t\ge 0$ in the Penrose diagram is depicted, and the dashed lines are the event horizons.
	We are interested only in the outside of those horizons.}
	\label{fig: multiple black holes}
\end{figure}

In this work,
we extend the framework \cite{Takeda:2024qbq} by introducing baths so that we can consider a generic notion of heat in black hole thermodynamics. 
This setup can be realized as follows.
We first prepare a target CFT and bath CFTs which are all assumed to be holographic.
Then, the target CFT is coupled to the bath CFTs by double-trace deformations \cite{Aharony:2001pa, Witten:2001ua}.
Finally, we consider the holographic dual of the whole system.
The bulk theory is given by multiple asymptotically AdS gravitational systems glued together along the boundary. 
They exchange heat through their asymptotic boundaries\footnote{
The setup of two coupled AdS theories is also studied recently in \cite{Bintanja:2023vel, Karch:2023wui, Geng:2023ynk} with different motivations.
} (Figure \ref{fig: multiple black holes}).
We explicitly show the process of taking the holographic dual using the path integral equivalence. Although the way to incorporate the double-trace deformation in AdS/CFT has already been developed in \cite{Aharony:2001pa, Witten:2001ua, Berkooz:2002ug, Sever:2002fk, Aharony:2005sh, Aharony:2006hz}, we provide a derivation in our setup.\footnote{
In the literature, \cite{Aharony:2001pa, Witten:2001ua, Berkooz:2002ug, Sever:2002fk, Aharony:2005sh} study the multi-trace deformation within one CFT, while \cite{Aharony:2006hz} suggests a way to incorporate the deformation involving different CFTs, providing some evidence.
We validate the latter suggestion from the path-integral equivalence.
}
The second law and other thermodynamic properties in the bulk are derived by translating those on the boundary into the bulk language.
Here, external sources applied to CFT operators, which are in the bulk the boundary values of the fields, play the role of work in the first law.

We also check our second law using a three-dimensional Einstein-scalar theory as an example.
We perturbatively calculate the entropy production around the BTZ black hole background.\footnote{Similar holographic computations are investigated in the context of quantum quench (see, e.g., \cite{Das:2010yw, Basu:2011ft, Buchel:2013lla, Das:2014jna, Berenstein:2014cia}).}
On the CFT side, we turn on a time-dependent source for a primary operator dual to the bulk scalar field.
For the simplicity of the calculation, we focus on $0<\Delta<2$ ($\Delta\neq 1$), where $\Delta$ is the conformal dimension of the boundary operator.
Then, we numerically find that the second law is satisfied for $0<\Delta<1$, while violated for $1<\Delta<2$.
Our second law is a consequence of the non-negativity of the relative entropy, and must not be violated at least if the theory is unitary. 
One might think that the violation is due to the invalidity of using the holographic duality for this bulk model.
However, this is not the case because our perturbative calculation only probes the universal sector, which is determined simply by the conformal symmetry shared by both the bulk and boundary theories.
Instead, we argue that the violation of the second law arises from the renormalization scheme, and that the renormalization scheme must be chosen carefully to avoid violating the property of the relative entropy.

This paper is organized as follows.
In section \ref{sec: boundary}, we begin with formulating the thermodynamics for generic composite quantum systems and derive their thermodynamic properties.
In section \ref{sec: bulk}, we focus on a system of a target CFT and a bath CFT, where both are assumed to be holographic. 
Based on the AdS/CFT dictionary, the system is shown to be dual to two AdS theories interacting via the asymptotic boundary conditions.
Then, the thermodynamic properties derived in section \ref{sec: boundary} are straightforwardly translated into the bulk black hole side.
This offers a formulation of thermodynamics for composite black hole systems.
In section \ref{sec: example}, we demonstrate the above results in two examples.
One is a single closed Einstein-scalar theory without a bath system, and another consists of two coupled Einstein-scalar theories.
The final section \ref{sec: summary} is devoted to summary and discussions.

In the main text, we focus on the simplest case where the composite system consists of two theories (a target system and a bath system).
In appendix \ref{app: multiple version}, we generalize the results in the main text to the multiple interacting theories.
In appendix \ref{app: example in CFT}, the calculations in section \ref{sec: example} are reproduced on the CFT side.
In appendix \ref{app: retarded}, we give a comment on the equivalence of the retarded Green's function in the boundary and the one computed in the bulk.

\section{Coarse-graining composite quantum system}\label{sec: boundary}
We consider a target system and a bath system interacting with each other as
\begin{align}\label{eq: QM Hamiltonian}
  H_\ttl(t) = H^\ith{s}(w(t)) + H^\ith{b}_* + V(t),
\end{align}
where $H^\ith{s}$ is the system Hamiltonian, $H^\ith{b}_*$ is the bath Hamiltonian independent of time, and $V(t)$ is the interaction part between them.
The system Hamiltonian can depend on time through protocol $w(t)$, which realizes the work done on the target system.
We set that $H_*^\ith{s}:= H^\ith{s}(0)$ corresponds to the time-independent Hamiltonian.
We also assume that the interaction between the system and the bath is turned off for $t<0$.
The initial state is set to be $\rho(0)$ at $t=0$, and the state evolves as
\begin{align}\label{eq: unitary evolution}
  \rho(t) = U(t) \rho(0) U(t)^\dagger,\qquad
  U(t) := \mathrm{T}\exp\l(-i\int_0^t\d t'\, H_\ttl(t')\r),
\end{align}
where $\mathrm{T}$ denotes the time-ordering symbol.
In section \ref{sec: boundary} to \ref{sec: example}, we only consider this simple case for ease of notations.
The generalization to the multiple interacting theories will be straightforwardly done in appendix \ref{app: multiple version}.

\subsection{Coarse-grained entropy}
\label{sec:cg-entropy}
We now construct the thermodynamics for the total system. 
First, we define the coarse-grained entropy. 
The entropy depends on the set of operators that we regard as thermodynamic observables. 
For simplicity, we here suppose that only $H_*^\ith{s}$, $O$, and $H^\ith{b}_*$ are available observables for us, where $O$ is a time-independent Schr\"odinger operator in the target system. 
The reason why we include $O$ in the thermodynamic observables is that we will associate the protocol $w(t)$ with $O$ and introduce the notion of work later.

At any time $t$, the expectation values of an operator $A$ are defined as
\begin{align}\label{eq: QM expectation value}
  \braket{A}_t := \Tr(\rho(t) A).
\end{align}

In thermodynamics, all we can do is to estimate the state from the accessible information.
Then, the principle of maximum entropy tells us that the most likely guess of the state is the maximum entropy state that respects the expectation values we are interested in.
Let $\bar\rho(t)$ be such a maximum entropy state at $t$.
We apply the Lagrange multiplier method to the von Neumann entropy under the following condition:
\begin{align}\label{eq: coarse-grained conditions}
   \Tr(\bar\rho(t) A)=\braket{A}_t,
   \qquad \mbox{for}\quad A = H_*^\ith{s,b},\, O.
\end{align}
One can easily find that
\begin{align}\label{eq: total coarse-grained state}
  &\bar\rho(t) = \bar\rho^\ith{s}(t)\otimes \bar\rho^\ith{b}(t),\nonumber\\
  &\bar\rho^\ith{s}(t) := \frac{1}{Z^\ith{s}(t)}e^{-\beta(t)(H_*^\ith{s}-\mu(t)O)},\qquad
  \bar\rho^\ith{b}(t) :=\frac{1}{Z^\ith{b}(t)}e^{-B(t)H^\ith{b}_*},
\end{align}
with each normalization factor given by the trace of the exponential as usual.
The multipliers $\beta(t)$, $\mu(t)$, and $B(t)$ are determined by \eqref{eq: coarse-grained conditions}.
We also call $\bar\rho(t)$ as coarse-grained state.
Note that $O$ does not necessarily commute with $H_*^\ith{s}$ in general.
Thermodynamics of non-commutative operators were introduced in \cite{PhysRev.108.171}, and a short historical review is summarized in section II of \cite{Majidy:2023}.

Using this coarse-grained state $\bar\rho(t)$, we can define the coarse-grained entropy $S^\ith{s}(t)$ and $S^\ith{b}(t)$ as
\begin{align}\label{eq: system bath entropy}
  S^\ith{s}(t) :=& -\Tr_s\l[\bar\rho^\ith{s}(t) \ln \bar\rho^\ith{s}(t)\r] = \beta(t)(\braket{H_*^\ith{s}}_t - \mu(t)\braket{O}_t) + \ln Z^\ith{s}(t),\nonumber\\
  S^\ith{b}(t) :=&-\Tr_b\l[\bar\rho^\ith{b}(t) \ln \bar\rho^\ith{b}(t)\r] = B(t)\braket{H^\ith{b}_*}_t + \ln Z^\ith{b}(t).
\end{align}
Here, we have used \eqref{eq: coarse-grained conditions} and $\Tr_{s}$ ($\Tr_{b}$) means the trace over the system (bath) degrees of freedom.
The total entropy is defined as the sum of them,
\begin{align}
  S(t) = S^\ith{s}(t) + S^\ith{b}(t).
\end{align}

\subsection{The second law}
As shown in \cite{Takeda:2024qbq}, the second law holds for the total system,
\begin{align}\label{eq: the total second law}
  S(t) \geq S(0).
\end{align}
This is derived from \eqref{eq: unitary evolution}, \eqref{eq: coarse-grained conditions}, and the non-negativity of the following relative entropy,
\begin{align}
  0 \leq S(\rho(t)||\bar\rho(t)):=\Tr\l[ \rho(t) (\ln \rho(t) - \ln \bar\rho(t))\r].
\end{align}
Here, we also need to assume that the initial state is of the form of the generalized Gibbs state, that is,
\begin{align}\label{eq: rho_0 condition}
  \rho(0) = \bar\rho(0).
\end{align}
This assumption \eqref{eq: rho_0 condition} corresponds to the situation where the system is set for $t<0$ in a steady state under the following Hamiltonian:
\begin{align}\label{eq: non-equiv Hamiltonian}
  H_{\mathrm{non-equiv}} = H_*^\ith{s} - \mu(0) O + H^\ith{b}_*.
\end{align}

The time evolution of $S^\ith{s}(t)$ can be calculated as follows.
Taking the time derivative of $S^\ith{s}(t)$, and using \eqref{eq: coarse-grained conditions}, \eqref{eq: total coarse-grained state} and \eqref{eq: system bath entropy}, we obtain
\begin{align}\label{eq: fundamental relation}
  \dot S^\ith{s}(t) = -\Tr_s\dot {\bar\rho}^\ith{s}(t) \ln \bar\rho^\ith{s}(t) -\Tr_s\dot {\bar\rho}^\ith{s}(t) = \beta(t)\left[\frac{d}{dt}\braket{H_*^\ith{s}}_t - \mu(t)\frac{d}{dt}\braket{O}_t \right].
\end{align}
In the last equality, we have used $\Tr_s\dot {\bar\rho}^\ith{s}(t) = 0$, which comes from $\Tr_s\bar\rho^\ith{s}(t)=1$.
This relation \eqref{eq: fundamental relation} is exactly the same as the fundamental thermodynamic relation.

Similarly, the fundamental relation for the bath is found as
\begin{align}\label{eq: bath fundamental relation}
  \dot S^\ith{b}(t) = B(t)\frac{d}{dt}\braket{H_*^\ith{b}}_t.
\end{align}
Now, the second law \eqref{eq: the total second law} can be written as
\begin{align}\label{eq: 2nd law with heat}
  S^\ith{s}(t) \geq S^\ith{s}(0) + \int_0^t\d t'\, B(t')\delta Q(t'),
  \qquad
  \delta Q(t) := - \frac{\d }{\d t} \braket{H_*^\ith{b}}_t.
\end{align}
This is exactly the same form as the second law of a target system attached to a bath.
In contrast to a single isolated system, the entropy of the target system $S^\ith{s}(t)$ can decrease due to the ``heat exchange" $\delta Q(t)$.

So far, we have considered only $H_*^\ith{s,b}$ and $O$ as operators to be respected.
In appendix \ref{app: multiple version}, we also consider the case of more operators, and derive the second law and the fundamental relation.

\subsection{The first law}\label{subsec: first law}
The first law of thermodynamics is a form of energy conservation law.
The notion of heat is understood as the energy flow that cannot be traced through mechanical operations.

The first law is commonly obtained as follows.
We first define the energy of the system by the time-dependent Hamiltonian:
\begin{align}\label{eq: internal energy}
  E(t) := \Tr_s\l[\bar\rho^\ith{s}(t) H^\ith{s}(w(t))\r].
\end{align}
Since the state of the total system is now represented by the coarse-grained state, we must use $\bar\rho^\ith{s}(t)$ instead of $\rho(t)$ to derive a thermodynamic law.
Also, we must use the time-dependent Hamiltonian $H^\ith{s}(w(t))$ instead of $H_*^\ith{s}$.
This is because the protocol $w(t)$ deforms the potential and thus energy levels, which should be counted as the energy portion of the target system,  as exemplified by the case where magnetic field is applied to a hydrogen atom.

Taking the time derivative of $E(t)$, we obtain
\begin{align}\label{eq: pre first law}
\dot E(t)
  = \Tr_s\l[\bar\rho^\ith{s}(t) \dot H^\ith{s}(w)\r] + \Tr_s\l[\dot {\bar\rho}^\ith{s}(t) H^\ith{s}(w)\r],
\end{align}
where the dot on $H^\ith{s}$ acts on the time-dependence of $w(t)$.
It is natural to regard the first term as the work, since it is the energy change driven by the protocol, which we can control by thermodynamic operations.
Thus, we name it as
\begin{align}\label{eq: work in QM}
  \delta W(t) := \Tr_s\l[\bar\rho^\ith{s}(t) \dot H^\ith{s}(w)\r] 
  = \dot w(t)~ \Tr_s\l[\bar\rho^\ith{s}(t) \frac{\partial H^\ith{s}(w)}{\partial w} \r].
\end{align}
This agrees with the common definition adopted in quantum thermodynamics.\footnote{
For example in the case of Lindbladian, the energy change from the time-dependence of the Hamiltonian is counted as work, and the rest part is counted as heat. See \cite{e15062100} as a review paper.
}
Then, the remaining term in \eqref{eq: pre first law}, which cannot be traced through mechanical thermodynamic operations, should be regarded as the heat:
\begin{align}\label{eq: tilde heat in QM}
  \delta \tilde Q(t) := \Tr_s\l[\dot {\bar\rho}^\ith{s}(t) H^\ith{s}(w)\r].
\end{align}
Finally we reach
\begin{align}\label{eq: 1st law}
  \dot E(t) = \delta W(t) + \delta \tilde Q(t).
\end{align}
We shall call this relation first law.

So far, we have obtained two definitions of heat, $\delta Q(t)$ in \eqref{eq: 2nd law with heat} and $\delta \tilde Q(t)$ in \eqref{eq: tilde heat in QM}.
The difference between them should be sufficiently small, because the heat in the first law must be the same as that in the second law.
To discuss the point, we here consider that $H^\ith{s}$ takes the form as $H^\ith{s} = H^\ith{s}_* + w(t) O$, which is of our interest later in section \ref{sec: bulk}.
Using \eqref{eq: coarse-grained conditions} and $\Tr(\dot \rho(t) H_\ttl) = 0$, which follows from the von Neumann equation, $\dot \rho(t) = i \cmt{\rho(t)}{H_\ttl}$, we find
\begin{align}\label{eq: delta heat difference}
    \delta Q(t) - \delta \tilde Q(t) = \Tr(\dot \rho(t) V) = i \Tr(\rho(t) \cmt{H^\ith{s}+H^\ith{b}_*}{V}).
\end{align}
To ensure that the difference can be ignored, it is sufficient to require that $\cmt{H^\ith{s}+H^\ith{b}_*}{V}$ almost vanishes as matrix elements.
Such an interaction is called resonant interaction \cite{potts2019introduction}.
The identification of $\delta Q(t)$ and $\delta \tilde Q(t)$ may be related to the emergence of the thermodynamics from microscopic theories.
Although it is a challenging and hot topic in physics, we in this paper do not pursue it furthermore.

Even in the more generic situation that we treat in appendix \ref{app: multiple version}, the first law \eqref{eq: 1st law} still applies as it is.

\section{Black hole thermodynamics via AdS/CFT}\label{sec: bulk}
In this section, we take the holographic dual of what we have constructed in the previous section.
Let $I^\ith{s}$ and $I^\ith{b}$ be the actions of $d$-dimensional holographic CFTs and $I_\ttl$ be the total action given by
\begin{align}\label{eq: CFT action}
  I_\ttl = I^\ith{s}[\gamma^\ith{s}] + I^\ith{b}[\gamma^\ith{b}] - \int\d^d x  \sqrt{-\gamma^\ith{s}}\, \l[w(x)O^\ith{s}(x) + v(x) O^\ith{s}(x)O^\ith{b}(x)\r].
\end{align}
Here, $x = (x^\mu)$ is the spacetime coordinate, $\gamma_{\mu\nu }^\ith{s}$ ($\gamma_{\mu\nu }^\ith{b}$) is the metric of the system (bath) theory, $O^\ith{s}(x)$ ($O^\ith{b}(x)$) is an operator in the system (bath) theory, and $w$ and $v$ are classical sources. 
The metrics $\gamma^\ith{s}$ and $\gamma^\ith{b}$ are distinguished at this stage to make the generating functional for the energy momentum tensors.
They are identified with each other at the end.
Even if one replaces $\gamma^\ith{s}$ in the interaction term with $\gamma^\ith{b}$, the equivalent result can be obtained.
The action \eqref{eq: CFT action} is comparable with \eqref{eq: QM Hamiltonian}. For example, the last term in \eqref{eq: CFT action} represents the interaction between the target system and the bath: it corresponds to the interaction $V(t)$ in \eqref{eq: QM Hamiltonian}.

The generating functional is formally expressed as
\begin{align}\label{eq: CFT Z}
  Z[\gamma,w,v] = \int \mathrm{D}\varphi\, e^{i I_\ttl[\varphi;\gamma,w,v]},
  \qquad
  \mathrm{D}\varphi := \mathrm{D}\varphi^\ith{s}\mathrm{D}\varphi^\ith{b},
\end{align}
where $\varphi^\ith{s}$ ($\varphi^\ith{b}$) denotes the elementary fields of the system (bath) theory collectively.
We here note that the symbol $\mathrm{D}$ indicates that the integral is performed for the fields defined on the boundary, because we will introduce another symbol $\mathbf{D}$ for bulk degrees of freedom below.
For more general setups (more theories and more operators), see appendix \ref{app: multiple version}.

\subsection{The holographic dual of system and bath}\label{subsec: taking dual}
The bulk gravitational theory that corresponds to the boundary theory \eqref{eq: CFT action} can be derived by the AdS/CFT dictionary.\footnote{We will consider only the classical limit of the bulk theory and its perturbative expansion with respect to the source $w(x)$ and the coupling $v(x)$ in \eqref{eq: CFT action}. 
Then, the use of the standard holographic dictionary can be justified at least in this regard.
As stated in \cite{Aharony:2006hz}, the stress tensor of each CFT is not conserved due to the interaction between the two systems, and it leads to a graviton mass in bulk: a linear combination of the two gravitons becomes massive. 
This generation of mass is a quantum effect in the bulk, and we can ignore it for our purpose.} We begin with \eqref{eq: CFT Z}.
First, by introducing auxiliary fields $\eta$ and $\chi$, the generating functional \eqref{eq: CFT Z} can be rewritten as
\begin{align}\label{eq: eta chi introduced}
 \int \mathrm{D}\varphi \mathrm{D}\eta \mathrm{D}\chi \exp\left[i I^\ith{s} + i I^\ith{b} + i \int\d^d x  \sqrt{-\gamma^\ith{s}}\left\{\eta  \left(vO^\ith{b}- \chi \right) -O^\ith{s}\left(w^\ith{1}+ \chi \right) \right\} \right].
\end{align}
Noticing that the integral over $\eta$ yields the delta function $\delta(\chi - vO^\ith{b})$, we see that \eqref{eq: eta chi introduced} goes back to \eqref{eq: CFT Z}.
Since the double-trace deformation is now resolved to the single-trace deformation, we can take the holographic dual for each theory as
\begin{align}\label{eq: taking dual}
  \int \mathbf{D}\Phi \mathrm{D}\eta \mathrm{D}\chi
  &
  \exp\left[i\mathcal{I}^\ith{s} + i\mathcal{I}^\ith{b} - i \int\d^d x  \sqrt{- \gamma^\ith{s}}\,\eta \chi \right]\nonumber\\
  \times
  &\delta\left(\hat \Phi^\ith{s} + w + \chi\right)
  \delta\left(\hat \Phi^\ith{b} - v\eta \right)
  \delta \left(\hat g^\ith{s}_{\mu\nu } - \gamma_{\mu\nu }^\ith{s} \right)
  \delta \left(\hat g^\ith{b}_{\mu\nu } - \gamma_{\mu\nu }^\ith{b} \right).
\end{align}
Here, $\mathcal{I}^\ith{s}$ ($\mathcal{I}^\ith{b}$) is the bulk action dual to the boundary action $I^\ith{s}$ ($I^\ith{b}$), and $\mathbf{D}$ denotes the integral over the bulk degrees of freedom.
The two theories $\mathcal{I}^\ith{s}, \mathcal{I}^\ith{b}$ live in different $(d+1)$-dimensional asymptotically AdS spaces, as suggested in \cite{Aharony:2006hz}.
The metric for $\mathcal{I}^\ith{s}$-theory is expressed in the Fefferman-Graham gauge as
\begin{align}
  G^\ith{s}_{MN}\d X^M \d X^N = \left(\frac{L^\ith{s}}{z} \right)^2\d z^2 + g^\ith{s}_{\mu\nu }\d x^\mu \d x^\nu,
\end{align}
with $L^\ith{s}$ being the AdS radius.
The metric for $\mathcal{I}^\ith{b}$-theory is also expressed similarly, but $g^\ith{b}_{\mu\nu}$ and the AdS radius $L^\ith{b}$ can differ from $g^\ith{s}_{\mu\nu}$ and $L^\ith{s}$ in general.
The symbol $\Phi$ collectively denotes the bulk metric fields $g^\ith{s,b}_{\mu\nu }$ and the matter fields $\Phi^\ith{s,b}$ dual to $O_J^\ith{s,b}$.
The hat quantities $\hat g^\ith{s}_{\mu\nu }$ and $\hat \Phi^\ith{s}$ are the renormalized boundary values such that,
\begin{align}
	g^\ith{s}_{\mu\nu } \sim \left(\frac{L^\ith{s}}{z} \right)^2\hat g^\ith{s}_{\mu\nu },\qquad
  \Phi^\ith{s} \sim z^{d-\Delta^\ith{s}}\hat \Phi^\ith{s}\qquad (z\to 0),
\end{align}
where $\Delta^\ith{s}$ is the conformal dimension of boundary operator $O^\ith{s}$.
We define $\hat g^\ith{b}_{\mu\nu }$ and $\hat \Phi^\ith{b}$ in a similar way.
Finally, performing the $\chi$-integral in \eqref{eq: taking dual}, we obtain
\begin{align}
   \int \mathbf{D}\Phi \mathrm{D}\eta
  &\exp\left[i \mathcal{I}^\ith{s} + i \mathcal{I}^\ith{b} + i \int\d^d x  \sqrt{- \gamma^\ith{s}}\,\eta \left(\hat \Phi^\ith{s} + w \right) \right]\nonumber\\
  &\times
  \delta\left(\hat \Phi^\ith{b} - \eta v \right)
  \delta \left(\hat g^\ith{s}_{\mu\nu } - \gamma_{\mu\nu }^\ith{s} \right)
  \delta \left(\hat g^\ith{b}_{\mu\nu } - \gamma_{\mu\nu }^\ith{b} \right).
\end{align}
We pose here since the $\eta$-integral cannot be performed in general as $v(x)$ can be zero.

In the large $N$ limit, we can evaluate the generating functional at the saddle point, which we assume is simply found by the action principle as usual.
The total action is now given by
\begin{align}\label{eq: bulk action}
\mathcal{I}_\ttl := \mathcal{I}^\ith{s} + \mathcal{I}^\ith{b} + \int\d^d x  \sqrt{- \gamma^\ith{s}}\,\eta \left(\hat \Phi^\ith{s} + w \right),
\end{align}
for which we solve the variational problem under the following conditions:
\begin{align}
  0 = \hat \Phi^\ith{b} - v\eta = \hat g^\ith{s}_{\mu\nu } - \gamma_{\mu\nu }^\ith{s}= \hat g^\ith{b}_{\mu\nu } - \gamma_{\mu\nu }^\ith{b}.
\end{align}
Then, the variations on the boundary are restricted as
\begin{align}
  0 = \delta \hat \Phi^\ith{b} -v\delta \eta = \delta \hat g^\ith{s}_{\mu\nu }= \delta \hat g^\ith{b}_{\mu\nu }.
\end{align}
Noting those conditions, we require
\begin{align}\label{eq: delta I gravity}
 0= \delta \mathcal{I}_\ttl &= (\mbox{EOMs}) + \int\d^d x\sqrt{-\gamma^\ith{s}}\left[\left(\Pi^\ith{s} + \eta \right)\delta \Phi^\ith{s} + \left(v \Pi^\ith{b} + \hat \Phi^\ith{s} + w \right)\delta \eta \right],\nonumber\\
 \mathrm{i.e.,}\quad
 0&=(\mbox{EOMs})=\Pi^\ith{s} + \eta = v \Pi^\ith{b} + \hat \Phi^\ith{s} + w,
\end{align}
with $\Pi^\ith{s,b}$ defined as
\begin{align}
  \Pi^\ith{s}(x) : =  \frac{1}{\sqrt{-\gamma^\ith{s}}}\frac{\delta \mathcal{I}^\ith{s}|_\mathrm{on-shell}}{\delta \hat \Phi^\ith{s}(x)},
  \qquad
  \Pi^\ith{b}(x) : =  \frac{1}{\sqrt{-\gamma^\ith{b}}}\frac{\delta \mathcal{I}^\ith{b}|_\mathrm{on-shell}}{\delta \hat \Phi^\ith{b}(x)}.
\end{align}
Therefore, by eliminating $\eta$, the remaining task is to find the solutions subject to
\begin{align}\label{eq: boundary conditions}
  0 = \hat \Phi^\ith{s} + w + v \Pi^\ith{b} =  \hat \Phi^\ith{b} + v \Pi^\ith{s} = \hat g^\ith{s}_{\mu\nu } - \gamma_{\mu\nu }^\ith{s}= \hat g^\ith{b}_{\mu\nu } - \gamma_{\mu\nu }^\ith{b},
\end{align}
and evaluate the generating functional with the on-shell bulk action,
\begin{align}\label{eq: bulk Z}
  -i \ln Z[\gamma,w,v] = \mathcal{I}_\ttl|_\mathrm{on-shell} = \mathcal{I}^\ith{s}+\mathcal{I}^\ith{b} + \int\d^d x  \sqrt{- \gamma^\ith{s}}\,v\Pi^\ith{s}\Pi^\ith{b}.
\end{align}
Note that the target system interacts with the bath system through the asymptotic boundary conditions while there is no interaction term explicitly in the EOMs.
The conditions \eqref{eq: boundary conditions} coincide with the ones developed before in \cite{Aharony:2001pa, Witten:2001ua, Berkooz:2002ug, Sever:2002fk, Aharony:2005sh, Aharony:2006hz}.
If $v$ is set to zero, the system reduces to the ordinary boundary value problem with the Dirichlet conditions.

At the end of this section, we derive the bulk expressions of the boundary one-point functions.
From \eqref{eq: CFT action}, \eqref{eq: CFT Z}, and the first equality of \eqref{eq: bulk Z}, we obtain
\begin{align}
  \braket{O^\ith{s}(x)} = \frac{i}{\sqrt{- \gamma^\ith{s}}} \frac{\delta \ln Z[\gamma ,w,v]}{\delta w(x)}
  = - \frac{1}{\sqrt{- \gamma^\ith{s}}} \frac{\delta \ln \mathcal{I}_\ttl|_\mathrm{on-shell}}{\delta w(x)}.
\end{align}
To evaluate this, it is convenient to recover the auxiliary field $\eta$ and use \eqref{eq: bulk action}.
Though the variation with respect to $w$ can affect all the bulk fields, owing to the EOMs and boundary conditions, only the variation of $w$ that explicitly appears in the last term of \eqref{eq: bulk action} survives.
Thus, we obtain
\begin{align}\label{eq: O expectation value}
  \braket{O^\ith{s}(x)} = -\eta(x) = \Pi^\ith{s}(x).
\end{align}
In the last equality, we have used \eqref{eq: delta I gravity}.

Similarly, taking the derivative with respect to $v$ and metrics, we obtain
\begin{align}
  &\braket{O^\ith{s} O^\ith{b}} = -\Pi^\ith{s} \Pi^\ith{b},
  \qquad
  \braket{T^\ith{b}_{\mu\nu }}
  =
  Y^\ith{b}_{\mu\nu },
  \label{eq: OO expectation value}\\
  &\braket{T^\ith{s}_{\mu\nu }} - \left(w\braket{O^\ith{s}} + v\braket{O^\ith{s} O^\ith{b}} \right)\gamma^\ith{s}_{\mu\nu }
  =
  Y^\ith{s}_{\mu\nu } + v\Pi^\ith{s}\Pi^\ith{b}\gamma^\ith{s}_{\mu\nu },
  \label{eq: EM expectation value}
\end{align}
where $Y^\ith{s,b}_{\mu\nu }$ is the (renormalized) Brown-York tensor \cite{Brown:1992br,Balasubramanian:1999re,deHaro:2000vlm},
\begin{align}\label{eq: Brown-York tensor}
  Y^\ith{s}_{\mu\nu }(x) := -\frac{2}{\sqrt{-\gamma^\ith{s}}} \frac{\delta \mathcal{I}^\ith{s}}{\delta \gamma_\ith{s}^{\mu\nu }(x)},\qquad
  Y^\ith{b}_{\mu\nu }(x) := -\frac{2}{\sqrt{-\gamma^\ith{b}}} \frac{\delta \mathcal{I}^\ith{b}}{\delta \gamma_\ith{b}^{\mu\nu }(x)}.
\end{align}
We see that the two-point function [the left equation in \eqref{eq: OO expectation value}] factorizes into the product of one-point functions \eqref{eq: O expectation value}, which happened due to the large $N$ limit.
Using \eqref{eq: O expectation value}, \eqref{eq: OO expectation value}, and \eqref{eq: EM expectation value}, we obtain\footnote{There is a caveat on this dictionary, which will be discussed in section \ref{sec: summary}. We skip this point here since it does not affect the following section \ref{sec: example}.}
\begin{align}\label{eq: EM expectation value extracted}
  \braket{T_{\mu\nu }^\ith{s}} = Y_{\mu\nu }^\ith{s} + w\Pi^\ith{s} \gamma_{\mu\nu }^\ith{s},\qquad
  \braket{T_{\mu\nu }^\ith{b}} = Y_{\mu\nu }^\ith{b}.
\end{align}

Since we have distinguished the metrics $\gamma^\ith{s}$ and $\gamma^\ith{b}$ only for deriving \eqref{eq: EM expectation value extracted}, we put $\gamma^\ith{s} =\gamma^\ith{b} =\gamma$ hereafter.
In particular, it is assumed that $\gamma$ is static and written as
\begin{align}\label{eq: boundary metric}
  \gamma_{\mu\nu }\d x^\mu \d x^\nu  = -\d t^2 + \sigma_{ab}\d x^a \d x^b,\qquad \sqrt{-\gamma} = \sqrt{\sigma}.
\end{align}
The indices $a, b, ...$ are used for the spacelike coordinates.
The results will be generalized to non-static cases straightforwardly.
On this metric, the Hamiltonian is given as
\begin{align}\label{eq: Hamiltonian and EM tensor}
    H^\ith{s}_* = \int d^{d-1}\vec x\sqrt{\sigma}T^\ith{s}_{00},\qquad
    H^\ith{b}_* = \int d^{d-1}\vec x\sqrt{\sigma}T^\ith{b}_{00}.
\end{align}

\subsection{Black hole thermodynamics with bath}\label{subsec: BHT}
So far, we have derived the bulk composite theory dual to composite boundary theory \eqref{eq: CFT action}.
Now, we need to specify the integration contour of the path integral \eqref{eq: CFT Z} in order to construct the black hole thermodynamics.
In \cite{Takeda:2024qbq}, where the target system is isolated, the author considered the Schwinger-Keldysh like integration contour based on the time evolution \eqref{eq: unitary evolution} with the initial condition \eqref{eq: rho_0 condition}, following the prescription given by \cite{Maldacena:2001kr,Skenderis:2008dg}.
The same strategy can be applied to the current case straightforwardly.
However, instead of repeating the same procedure, we here give a more intuitive explanation to identify the bulk initial value problem which is dual to the CFT evolution \eqref{eq: unitary evolution}.

We respect an operator $O^\ith{s}(x)$ and the Hamiltonians $H_*^\ith{s}$ and $H_*^\ith{b}$ as in section~\ref{sec: boundary}.\footnote{
In appendix \ref{app: multiple version}, we will show that momentum operators can also be included in the coarse-graining. This enables us to respect the (angular) momenta of the bulk black hole.
}
The maximum entropy state we prepare at $t = 0$ is expressed as $\rho(0) = \rho^\ith{s}(0)\otimes \rho^\ith{b}(0)$ with
\begin{align}
  \rho^\ith{s}(0) &\propto \exp\left[-\beta(0) \left(H^\ith{s}_* - \int \d^{d-1} \vec x\sqrt{\sigma} \mu(0,\vec x)O^\ith{s}(\vec x) \right) \right],\quad
  \rho^\ith{b}(0) &\propto e^{-B(0)H^\ith{b}_*}.
\end{align}
Note again that $O^\ith{s}(\vec x)$ is in the Schr\"odinger picture.

The system evolves toward the future with the time-dependent Hamiltonian specified by the action \eqref{eq: CFT action}, that is,
\begin{align}
  H_\ttl = H_*^\ith{s} + H_*^\ith{b} + \int\d^{d-1}\vec{x} \sqrt{\sigma} \left[w(t,\vec x)O^\ith{s}(\vec x)+ v(t,\vec x)O^\ith{s}(\vec x) O^\ith{b}(\vec{x}) \right]\quad (t>0).
\end{align}
In order to obtain the bulk picture, we would like to extend the time evolution also toward the past.
Since we are interested in the thermodynamics of the system evolving from the steady state $\rho(0)$, we would like to keep the state in $\rho(0)$ even for the past $t<0$.
For $\rho(0)$ to be stationary, we must choose the Hamiltonian for $t<0$ as
\begin{align}\label{eq: t<0 Hamiltonian}
  H_\ttl = H_*^\ith{s} + H_*^\ith{b} - \int \d^{d-1} \vec x\sqrt{\sigma} \mu(0,\vec x) O^\ith{s}(\vec x)\qquad (t<0),
\end{align}
using \eqref{eq: non-equiv Hamiltonian}.
The system and the bath are completely decoupled for $t<0$.

\begin{figure}
	\centering
	\includegraphics[width=0.9\columnwidth]{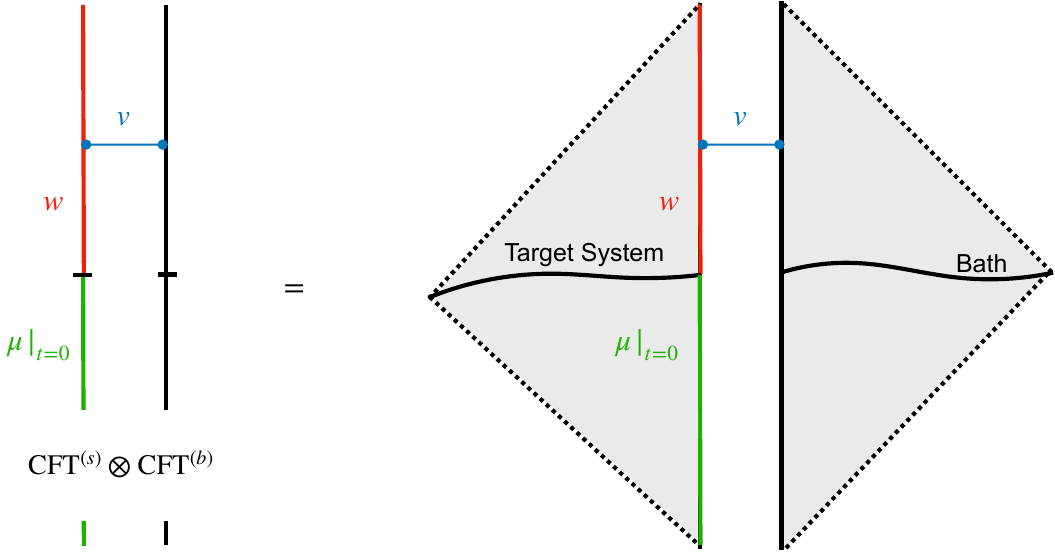}
	\caption{The system and baths on both boundary and bulk theories.
	We construct the thermodynamics for the multiple AdS black holes outside the horizons.}
	\label{fig: duality}
\end{figure}

Therefore, the bulk dual of this CFT evolution would be constructed as follows (figure~\ref{fig: duality}).
We arbitrarily extend the time foliation to the bifurcation surface of the horizon in each spacetime, and the bulk region $t<0$ is determined so that it is dual to the stationary state $\rho(0) = \rho^\ith{s}(0)\otimes \rho^\ith{b}(0)$.
In other words, the solution is stationary and satisfies the asymptotic boundary conditions specified as
\begin{align}\label{eq: Lorentzian past BC}
  \hat \Phi^\ith{s}(t, \vec x) = \mu^\ith{s}(0,\vec x),\qquad
  \hat \Phi^\ith{b}(t,\vec x) = 0\qquad
  (t<0).
\end{align}
Having made the spacetime region $t<0$, we now have the configuration of the bulk fields and their conjugate momenta at the time slice $t = 0$.
Thus, the region $t> 0$ is determined by solving the initial value problem from $t = 0$ with the asymptotic boundary conditions \eqref{eq: boundary conditions} with $\gamma^\ith{s,b} = \gamma$ and \eqref{eq: boundary metric}.
The solution does not depend on how to specify the initial surface $t=0$ in the bulk, because all the possible initial surfaces share the same domain of dependence.

The bulk constructed in this way is practically equivalent to the one derived by following \cite{Takeda:2024qbq,Maldacena:2001kr,Skenderis:2008dg}.
Below, we consider the thermodynamics for this composite AdS black hole system.

\subsubsection*{Coarse-graining the dynamical black holes}
We establish the bulk description of the coarse-graining that corresponds to what we have done in section \ref{sec: boundary}.
The coarse-grained state of $\rho(t)$, which is denoted as $\bar\rho(t)$, is introduced as \eqref{eq: total coarse-grained state}.
Now for the CFTs, the coarse-grained state \eqref{eq: total coarse-grained state} is expressed with
\begin{align}
  \bar\rho^\ith{s}(t) &\propto \exp\left[-\beta(t) \left(H^\ith{s}_* - \int \d^{d-1} \vec x\sqrt{\sigma} \mu(t,\vec x) O^\ith{s}(\vec x) \right) \right],\\
  \bar\rho^\ith{b}(t) &\propto e^{-B(t)H^\ith{b}_*},
\end{align}
where $\beta(t)$, $B(t)$, and $\mu(t,\vec x)$ are the Lagrange multipliers.
The normalization factors, that is, the partition functions, are given as
\begin{align}\label{eq: CFT partition function}
  Z^\ith{s}(t) &=\Tr_s \exp\left[-\beta(t) \left(H^\ith{s}_* -  \int \d^{d-1} \vec x\sqrt{\sigma} \mu(t,\vec x) O^\ith{s}(\vec x) \right) \right],\\
  Z^\ith{b}(t) &= \Tr_b\, e^{-B(t)H^\ith{b}_*}.
\end{align}

The bulk description of $Z^\ith{s}(t)$ is a stationary Euclidean solution that matches the following asymptotic boundary conditions:
\begin{align}\label{eq: Euclidean BC}
  \hat \Phi^\ith{s}(\tau,\vec x) = -\mu(t,\vec x),\quad
  \hat g^\ith{s}_{\mu \nu }\d x^\mu \d x^\nu  = \d \tau^2 + \sigma_{ab} dx^a dx^b.
\end{align}
This is also explained in \cite{Takeda:2024qbq} by rewriting $Z^\ith{s}(t)$ in the Euclidean path integral and applying the AdS/CFT dictionary.
Note that $\tau$ is the Euclidean time coordinate whose periodicity is $ \beta(t)$ and is independent of the real-time $t$ (carefully check the arguments in \eqref{eq: Euclidean BC}).
Note also that the Euclidean solution is not directly related to the real-time bulk illustrated in figure~\ref{fig: duality}: the real-time bulk is dual to $\rho(t)$, and for each $t$ we consider the coarse-grained state $\bar{\rho}$ to which the stationary Euclidean solution is dual.
Similarly, the bulk description of $Z^\ith{b}(t)$ is simply the static Euclidean pure gravity solution that is not charged and rotating. 
If $\mathcal{I}^\ith{b}$ is the Einstein gravity, it is nothing but the Euclidean Schwarzschild-AdS with temperature $B(t)$.

So far, we have not yet discussed how to determine the multipliers within the bulk language.
In the boundary theory, they are fixed so that the expectation values of $O^\ith{s}$, $H^\ith{s}_*$ and $H^\ith{b}_*$ at each time $t$ are reproduced by the coarse-grained state.
As we have the bulk expressions of the expectation values owing to \eqref{eq: O expectation value}, \eqref{eq: EM expectation value extracted}, \eqref{eq: Hamiltonian and EM tensor}, and their Euclidean counterparts, the process of finding the multipliers can also be performed entirely within the bulk picture.
In particular, the expectation values of the Hamiltonian correspond to the mass of the bulk black holes, both in Lorentzian and Euclidean signatures.

\subsubsection*{The coarse-grained entropy and the second law}
The coarse-grained entropy can be rewritten by $\beta(t)$-derivative as\footnote{Strictly speaking, this must be seen as differentiating the following after which the proper values are substituted:
$$Z^\ith{s}[\beta, \mu] := \Tr_s \exp\left[-\beta \left(H^\ith{s}_* - \int \d^{d-1} \vec x\sqrt{\sigma} \mu O^\ith{s} \right) \right].
$$
}
\begin{align}\label{eq: entropy in CFT}
  S^\ith{s}(t) = -\left(\beta(t) \right)^2\frac{ \partial}{\partial \beta(t)}\left[\left(\beta(t) \right)^{-1} \ln Z^\ith{s}(t) \right].
\end{align}
Since $Z^\ith{s}(t)$ is dual to the on-shell bulk action evaluated by an Euclidean solution, this is also understood entirely in terms of the bulk gravity.
As in \cite{Takeda:2024qbq}, if $\mathcal{I}^\ith{s}$ is the Einstein gravity, \eqref{eq: entropy in CFT} is identified as the horizon area $A^\ith{s}(t)$ of the Euclidean spacetime, i.e.,
\begin{align}\label{eq: bulk entropy}
  S^\ith{s}(t) = \frac{A^\ith{s}(t)}{4G^\ith{s}},
\end{align}
with $G^\ith{s}$ being the Newton constant of $\mathcal{I}^\ith{s}$.\footnote{If not negligible, the matter entropy must be taken into consideration as the quantum correction.}
The same statement also holds for the bath.
Therefore, the second law \eqref{eq: the total second law} says,
\begin{align}\label{eq: bulk 2nd law}
  \frac{A^\ith{s}(t)}{4G^\ith{s}} + \frac{A^\ith{b}(t)}{4G^\ith{b}} \geq \frac{A^\ith{s}(0)}{4G^\ith{s}} + \frac{A^\ith{b}(0)}{4G^\ith{b}}.
\end{align}
Transforming this to \eqref{eq: 2nd law with heat}, we obtain another equivalent form,
\begin{align}
 \varsigma_t \geq 0,
\end{align}
where $\varsigma_t$ is the entropy production for the system black hole,
\begin{align}
  \varsigma(t) &:= \frac{A^\ith{s}(t)}{4G^\ith{s}} - \frac{A^\ith{s}(0)}{4G^\ith{s}} + \int_0^t\d t'\,B(t')\dot M^\ith{b}(t),\\
  M^\ith{b}(t) &:= \int \d^{d-1}\vec x \sqrt{\sigma}\, Y^\ith{b}_{00}.\label{eq: bath mass}
\end{align}
This follows from \eqref{eq: EM expectation value extracted} and \eqref{eq: Hamiltonian and EM tensor}.
Note that as opposed to the literature, we did not require any energy condition in deriving the second law of black hole thermodynamics, but consulted the holographic dictionary instead.
Our second law comes from the non-negativity of the relative entropy on the CFT side, and hence it must hold in any well-defined quantum theories.
Thus, the bulk inequality \eqref{eq: bulk 2nd law} (or generalized one using the Wald entropy) can serve as a criterion for whether a gravitational model of interest is in the landscape or the swampland.

\subsubsection*{The fundamental thermodynamic relation and the first law}
Although we dare not write it down explicitly, the fundamental relation \eqref{eq: fundamental relation} and the first law \eqref{eq: 1st law} can also be translated to the gravity side.
For the fundamental relation, we have already known how to calculate the entropy and the expectation values in \eqref{eq: fundamental relation} by the bulk language.
On the other hand, to reproduce the first law \eqref{eq: 1st law}, we need the counterparts of $\delta W(t)$ and $\delta \tilde Q(t)$.
The former is easier, since it is given as
\begin{align}\label{eq: bulk delta W}
  \delta W(t) = \int\d ^{d-1}\vec x\sqrt{\sigma}\,\dot w(t,\vec x) \Pi^\ith{s}(t,\vec x),
\end{align}
by \eqref{eq: QM Hamiltonian}, \eqref{eq: CFT action}, and \eqref{eq: O expectation value}.
The heat $\delta \tilde Q(t)$ is obtained indirectly by evaluating \eqref{eq: internal energy} in the bulk, taking its time derivative, and subtracting \eqref{eq: bulk delta W} from it.
Thus, as the ordinary thermodynamics, the heat is here again treated as the energy missing from the conservation law, which can only be determined after calculating the traceable energy.
In the situation that we discussed around \eqref{eq: delta heat difference}, $\delta \tilde Q(t)$ can be approximated by the energy change of the bath, which is nothing but $M^\ith{b}(t) - M^\ith{b}(0)$.

\section{Einstein-scalar examples}\label{sec: example}
We investigate the second law in the three-dimensional Einstein-scalar theory as an example. As mentioned in section~\ref{sec:intro}, there seems to be the case that the second law is violated. 
We argue that this is owing to the conventional holographic renormalization scheme. 
In section \ref{subsec: single example}, the system is isolated, and in section \ref{subsec: double example}, the target system interacts with a bath system.

\subsection{Single system}\label{subsec: single example}
We consider the bulk dual of the following boundary theory on the two-dimensional flat spacetime: 
\begin{align}\label{eq: example Hamiltonian}
  H(t) = H_* + \int \d \theta\, w(t) O(\theta),
\end{align}
where $H_*$ is the Hamiltonian of a holographic CFT, and $O$ is a primary scalar operator with conformal dimension $\Delta$. 
In this example, we have no bath system interacting with this target system. 
We suppose that the bulk theory dual to $H_*$ is the Einstein gravity and $O$ is dual to a bulk scalar field $\Phi$.
The protocol $w(t)$ is assumed to be uniform in the spatial direction ($\theta$-direction) for simplicity, although it can depend on $\theta$ in general.
Because of this, the computation below does not change even if the $\theta$-direction is compactified.
In the following, the protocol $w(t)$ is understood as a Schwartz function whose support is $[0,\infty)$.

Accordingly, the bulk theory that we consider is given as
\begin{align}
  &\mathcal{I}= \frac{1}{2\kappa}\int_M \d^3X \sqrt{-G}\left[R + \frac{2}{L^2} - C\partial_\mu \Phi\partial^\mu \Phi + C\frac{\Delta(2-\Delta)}{L^2}\Phi^2 \right]\nonumber\\
  &\hspace{180pt} +\frac{1}{\kappa}\int_{\partial M} \d^2 x\sqrt{-g}\left[K - \frac{1}{L} - C \frac{2-\Delta}{2L}\Phi^2 \right],\label{eq: example action} \\
  &G_{MN}|_{r=L/\epsilon} \d X^M \d X^N = \epsilon^2 \d r^2 + g_{\mu\nu }\d x^\mu \d x^\nu,\qquad
  g_{\mu\nu }\d x^\mu \d x^\nu =  \epsilon^{-2}\left(-\d t^2 + \d\theta^2 \right),\label{eq: G BC}
  \\
  &\Phi|_{r=L/\epsilon}= - (L\epsilon)^{2-\Delta}w(t),\label{eq: Phi BC}
\end{align}
where $R$ is the Ricci curvature, $L$ is the AdS radius, $C$ is some positive constant introduced to represent an ambiguity of the normalization of $\Phi$, and $\epsilon$ is the UV cutoff.
We have included the counterterms in $\mathcal{I}$.
The dimension $\Delta$ is chosen as $0<\Delta <2$ ($\Delta\neq 1$) for the simplicity of the counterterms \cite{deHaro:2000vlm}.
In this model, the mass dimensions of the parameters and fields are $[\kappa] = [L] = -1$, $[\Phi] = 0$, and $[w] = 2-\Delta$.

Our task is to compute
the following expectation values by finding the classical solution and using \eqref{eq: O expectation value} and \eqref{eq: EM expectation value} as
\begin{align}\label{eq: example Pi}
  \Pi(x) &= -\frac{CL}{\kappa} (L\epsilon)^{-\Delta} \left((2-\Delta)\Phi + r\partial_r \Phi \right)|_{r=L/\epsilon} \underset{\mathrm{dual}}{=} \braket{O(\theta)}_t,\\
  Y_{\mu\nu }(x) &= \frac{1}{\kappa}\left[\left(K - \frac{1}{L}- C\frac{2-\Delta}{2L}\Phi^2 \right)\gamma_{\mu\nu } - K_{\mu\nu }\right]_{r=L/\epsilon}\underset{\mathrm{dual}}{=} \braket{T_{\mu\nu }(\theta) - w(t)O(\theta)\gamma_{\mu\nu }}_t,\label{eq: example EM}
\end{align}
where $\gamma_{\mu\nu }$ is the renormalized boundary metric ($\gamma_{\mu\nu }\d x^\mu \d x^\nu = -\d t^2 + \d \theta^2$), $K_{\mu\nu }$ is the outward extrinsic curvature of the boundary, and $K$ is its trace $K := \gamma^{\mu\nu }K_{\mu\nu }$.
The classical bulk configuration is the $s$-wave solution since they are excited only through the boundary condition \eqref{eq: Phi BC} and $w(t)$ is uniform in the $\theta$-direction.
Thus, $\Pi$ and $Y_{\mu\nu }$ are independent of $\theta$.

\subsubsection*{Lorentzian evolution}
Now we solve the bulk theory perturbatively assuming that the source $w(t)$ is small as $w = \cO(\lambda)$ where $\lambda$ is a dimensionless small parameter $(\lambda \ll 1)$.
We will give the solution to $\cO(\lambda^2)$ which involves the back reaction from the scalar $\Phi$. 
Indeed, we need to consider $\cO(\lambda^2)$ to obtain a non-vanishing entropy production and have to go beyond the linear perturbation.
An equivalent computation on the CFT side is also investigated in appendix~\ref{app: example in CFT}.

The $\cO(\lambda^0)$ solution is the BTZ black hole with radius $r_+$,
\begin{align}
\label{BTZ0}
  \d s^2 = \frac{\d r^2}{N^2} - N^2 \d t^2 + \frac{r^2}{L^2} \d \theta^2,\qquad N^2 := \frac{r^2 - r_+^2}{L^2}.
\end{align}
This is a vacuum solution and $\Phi \equiv 0$ in this order.

To obtain the leading $\cO(\lambda)$ solution with $C = \cO(\lambda^0)$, we solve the Klein-Gordon equation on the above BTZ background.
As explained around the figure \ref{fig: duality}, the bulk is stationary for the past $t<0$.
In the current case, the bulk is the BTZ background with $\Phi\equiv 0$.
The bulk configuration deviates from the stationary state only after $t=0$, when $\Phi$ begins to be excited by the source $w(t)$.
The solution $\Phi$ that satisfies this initial condition and the condition \eqref{eq: Phi BC} is found to be
\begin{align}
  &\Phi(t,r) = - \int \d \omega\, e^{-i\omega t}\, w_\omega \left(\frac{L^2}{r } \right)^{2-\Delta}\left(1-\frac{r_+^2}{r^2} \right)^{-\frac{i\omega L^2}{2r_+}}\left(1-\frac{r_+^2\epsilon^2}{L^2} \right)^{\frac{i\omega L^2}{2r_+}}\frac{F_\omega (r)}{F_\omega (L/\epsilon)} ,\label{eq: Phi solution} \\
  &\alpha_\omega := \frac{\Delta}{2} -\frac{i\omega L^2}{2r_+},\quad
  \gamma_\omega := 1 -\frac{i\omega L^2}{r_+},\quad
  F_\omega (r) := \._2F_1\left(\gamma_\omega - \alpha_\omega, \gamma_\omega - \alpha_\omega,\gamma_\omega; 1- \frac{r_+^2}{r^2} \right),\\
  &w_\omega := \int_{-\infty}^\infty \frac{\d t}{2\pi} e^{i\omega t}w(t) = \int_0^\infty \frac{\d t}{2\pi} e^{i\omega t}w(t),\label{eq: w FT}
\end{align}
where $\._2F_1$ is the hypergeometric function.
One can easily check that this solution satisfies the asymptotic boundary condition \eqref{eq: Phi BC} appropriately.
The easiest way to confirm the initial condition, $\Phi\equiv 0$ ($t<0$), is to notice that the wave is purely incoming near the horizon.
If there were an outgoing mode, that mode would have come from the past.
Of course, the initial condition can also be directly shown from the fact that the $\omega$-integration contour in \eqref{eq: Phi solution} can be deformed away to the semicircle at the infinity on the upper half plane, for $t<0$.
Here, note from \eqref{eq: w FT} that $w_\omega$ is regular everywhere on the upper half plane and rapidly goes to zero at infinity.
Therefore, from \eqref{eq: example Pi} and \eqref{eq: Phi solution}, we obtain by taking the limit $\epsilon\to 0$,
\begin{align}\label{eq: final Pi}
  \Pi(t) = \frac{2CL\,\Gamma(2-\Delta)^2}{\pi \kappa}\left(\frac{r_+}{L^2} \right)^{2\Delta-2}(-\sin(\pi \Delta)) \int\d \omega\, e^{-i \omega t}w_\omega \left(\frac{\Gamma(\alpha_\omega)}{\Gamma(\gamma_\omega - \alpha_\omega)} \right)^2.
\end{align}

Next, we compute the back reaction to the metric due to the $\cO(\lambda)$ matter \eqref{eq: Phi solution}. The back reaction is $\cO(\lambda^2)$.
Let $h_{MN}$ be the perturbed part from the BTZ background, and we gauge-fix it as \cite{Rocha:2011wp},
\begin{align}
  h_{MN}\d X^M \d X^N = \frac{A(r,t)}{N^4}\d r^2 + B(r,t)\d t^2 + 2 W(r,t)\d t \d \theta.
\end{align}
We now solve the perturbative Einstein equation, following the calculations in \cite{Rocha:2011wp}.
As in the case of $\cO(\lambda)$, perturbations are not turned on for the past $t<0$.
The perturbed part $h_{MN}$ contains only the normalizable mode, since the BTZ metric exactly satisfies the asymptotic boundary condition \eqref{eq: G BC}.
Although the exact solution can be found, here we just show the results that are necessary to calculate the energy: as $r \to \infty$,
\begin{align}
  &\frac{A-B}{N^2} \sim 0\\
  &A\sim -C(2-\Delta)(w(t)L^{2-\Delta})^2\left(\frac{r}{L} \right)^{2\Delta-2}\nonumber\\
  &\hspace{60pt} + \frac{\kappa L}{\Delta -1} \int_0^t \d t'\left[\Delta \dot w(t') \Pi(t') + (2-\Delta) w(t') \dot \Pi(t') \right].
\end{align}
Additionally, the off-diagonal component $W(r,t)$ vanishes everywhere.
Then, the mass density is computed as
\begin{align}
  m(t) := Y_{tt} = \frac{r_+^2}{2 \kappa L^3} + \int_0^t \d t'\, \dot w(t') \Pi (t').
\end{align}
Extracting the source term and using $w(0)=0$, we obtain
\begin{align}\label{eq: h_t}
  h(t) := m(t) - w(t) \Pi(t) = \frac{r_+^2}{2 \kappa L^3} - \int_0^t \d t'\, w(t')\dot \Pi (t'),
\end{align}
which is dual to $\braket{H_*}_t$.\footnote{\label{foot:mass_density}
More precisely, $h(t)$ is a density and dual to $\braket{H_*}_t/L_\theta$ where $L_\theta$ is the volume of the space where the CFT lives, i.e., $L_\theta$ is the length for the $\theta$-direction.}

\subsubsection*{Coarse-graining the Lorentzian black hole with Euclidean solution}
We build the coarse-grained Euclidean black hole by respecting only the Hamiltonian $H_\ast$, i.e., the above $h(t)$ in \eqref{eq: h_t}. 
We do not respect $O$ here because of the following reasons.
As can be checked by a similar calculation above, we cannot find the static and rotationally symmetric Euclidean solution with scalar hair, at least perturbatively to the order of our interest.
Originally, the action \eqref{eq: example action} may contain higher order interaction terms, although we have ignored them because they do not contribute in our perturbative order.
This means that, in the current example, $Z^\ith{s,b}(t)$ in \eqref{eq: CFT partition function} cannot be approximated by a simple saddle point, or that there is no perturbative solution for the Lagrange multipliers that respect both $H_*$ and $O$ at the same time.

The Euclidean BTZ spacetime with mass density $h(t)$ in \eqref{eq: h_t} is easily found, by keeping $\cO(\lambda^2)$ terms, as
\begin{align}
  \d s^2 =& \frac{\d r^2}{N_t^2} + N_t^2 \d \tau^2 + \frac{r^2}{L^2} \d \theta^2,\qquad
  N_t^2 := \frac{r^2 - (2\pi L^2T(t))^2}{L^2},\\
  T(t) :=& \frac{r_+}{2\pi L^2} - \frac{\kappa L}{2\pi r_+}\int^t_0\d t'\,w(t') \dot \Pi(t').
\end{align}
Thus, the entropy density\footnote{We have the translation symmetry in $\theta$-direction and thus the area of the horizon is proportional to the length of $\theta$-direction $L_\theta$ as in footnote~\ref{foot:mass_density}. The entropy density $s$ is the entropy divided by $L_\theta$.} is obtained from \eqref{eq: bulk entropy} with $\kappa=8\pi G$ as
\begin{align}\label{eq: s}
  s(t) = \frac{2\pi}{\kappa}\times 2\pi L^2 T(t)  = \frac{2 \pi r_+}{\kappa} - \frac{2\pi L^3}{r_+} \int_0^t \d t'\, w(t') \dot \Pi(t').
\end{align}
The second law, $s(t) \geq s(0)$, claims that the following inequality must hold:
\begin{align}\label{eq: single second law}
    \forall\, t>0,\quad
    -\int_0^t \d t'\, w(t') \dot \Pi(t')\geq 0.
\end{align}
Note that while the first term in \eqref{eq: s} is the entropy density for $\mathcal{O}(\lambda^{0})$ solution \eqref{BTZ0}, the last term is the correction due to the perturbation $w$ and is $\mathcal{O}(\lambda^2)$.\footnote{The two terms in \eqref{eq: s} are the same order with respect to $\kappa$ because we take the normalization $C=\mathcal{O}(\kappa^0)$ and $w=\mathcal{O}(\kappa^0)$, or $C=\mathcal{O}(\kappa)$ and $w=\mathcal{O}(\kappa^{-1/2})$. 
It depends on whether $w$ is intensive or extensive. 
Here we suppose that $\lambda$ and $\kappa$ are independent parameters although they can be correlated, for example, by taking $\lambda \sim 1/c$ where $c$ is the central charge of our holographic CFT.}
The absence of $\cO(\lambda)$ correction can be easily understood from the CFT computations as done in appendix~\ref{app: example in CFT} and is related to the non-negativity of the relative entropy.
This implies that the second law requires the absence of $\cO(\lambda)$ correction. 
Indeed, if the leading correction is linear in the source $w$, we can change the sign of the correction by flipping the sign of $w$  and it leads to the violation of the second law. 

The above result \eqref{eq: s} is universal in the sense that it is independent of the details of our holographic CFT as follows.
First, it is independent of the length of the $\theta$-direction and also does not depend on whether the $\theta$-direction is compactified or not because we take a spatially uniform source $w(t)$, as mentioned below \eqref{eq: example Hamiltonian}.
The entropy production (the last term in \eqref{eq: s}) is related to the back reaction due to the perturbation by $O$ and is related to the CFT three-point function $\braket{OTO}$ ($T$ is the CFT stress tensor).
Then, this three-point function is universal on the infinite line at the finite temperature, as is the case for the two-point function $\braket{OO}$. 
Indeed, we can obtain the same result [see \eqref{delEeq} with \eqref{S=betaE}] in general two-dimensional CFTs (not restricted to the holographic CFT).
Of course, if we take a non-uniform source $w(t,\theta)$ and the $\theta$-direction is periodic, the result depends on the details of CFTs because $\braket{OO}$ depends on them.

\subsubsection*{The second law, renormalization, and non-negativity of relative entropy}
\begin{figure}
    \centering
    \begin{minipage}{0.45\columnwidth}
        \centering
        \includegraphics[width = \columnwidth]{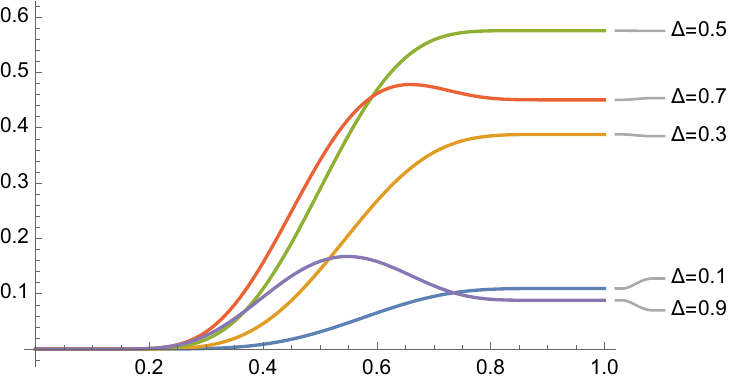}  
    \end{minipage}
    \hspace{12pt}
    \begin{minipage}{0.45\columnwidth}
        \centering
        \includegraphics[width = \columnwidth]{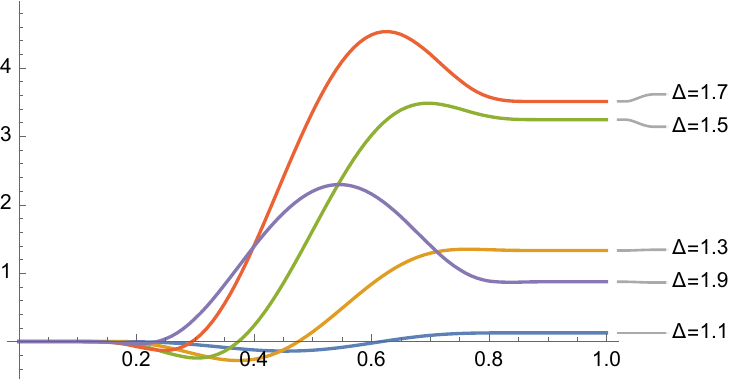}  
    \end{minipage}    
    \caption{The numerical check of the second law. The horizontal axis is the time and the vertical axis is the l.h.s.\ of \eqref{eq: single second law} up to a positive overall factor.
    We set $r_+/L^2 = 1$ here.
    The second law states that they should all be positive for all $t>0$.
    }
    \label{fig: bulk second}
\end{figure}

Figure \ref{fig: bulk second} shows our numerical check of the second law \eqref{eq: single second law} for some $\Delta$ with a source
\begin{align}
    w(t) = 
    \left\{
    \begin{array}{cc}
        \lambda\, \exp\left(-\frac{1}{t}-\frac{1}{1-t} + 4 \right) & (0< t<1) \\
        0 & (\mbox{otherwise})
    \end{array}
    \right..
\end{align}
Here we have chosen the form of $w(t)$ so that it is a Schwartz function, which is $C^\infty$ everywhere.
Figure~\ref{fig: bulk second} indicates that the second law is satisfied for all $t$ when $0<\Delta<1$.
However, the second law is violated for early $t$ when $1<\Delta<2$.
In the following, we discuss why in the latter case the second law is violated.

First of all, we discuss the CFT side.
The same quantity can be computed purely in CFT as mentioned above.
The one-point function is given by the retarded Green's function $G_R(t,\theta)$ as
\begin{align}\label{eq: CFT one point}
    \braket{O}_t = \int_{-\infty}^\infty\d t' \int\d \theta'\,G_R(t-t',\theta') w(t').
\end{align}
See \eqref{eq: CFT retarded function}, \eqref{eq: Pauli-Jordan} and \eqref{Guv} for the explicit form of $G_R$.
Since $G_R$ is singular when the two arguments are null-separated, we need to introduce a regulator ($i\epsilon$-prescription) as in \eqref{Guv} and send it to zero after performing the integral of the r.h.s.\ of \eqref{eq: CFT one point}.
Then on the CFT side, the entropy production is given as \eqref{eq: s} with $\Pi(t)$ replaced by $\braket{O}_t$ [see \eqref{def:del-calE}].
However, $\braket{O}_t$ diverges for $\Delta>1$ at any time $t$ when the source is turned on ($w(t)\neq 0$).
Therefore, we need to keep the regulator finite in that case.
While the entropy then depends on the regulator, the second law holds in CFT as desired by keeping the finite regulator (see appendix \ref{app: example in CFT}).

Then, what is happening on the bulk side?
We have obtained \eqref{eq: final Pi}.
Note that it is finite even for $\Delta>1$ (recall that $w(t)$ is a Schwartz function and so is its Fourier transform by definition).
This tells us that the bulk expression $\Pi(t)$ is renormalized in some way.
Indeed, we added the counterterms in \eqref{eq: example action} that are the standard ones in the holographic renormalization \cite{deHaro:2000vlm}, and these terms lead to the finite $\Pi(t)$, that is, a renormalized one-point function of $O$. 
This holographic renormalization can be understood as the analytic continuation of $\Delta$.
For $0<\Delta <1$, $\braket{O}_t$ in CFT is finite [see \eqref{delO_CFT_omega}] and it agrees with the bulk result $\Pi(t)$ in \eqref{eq: final Pi} (see appendix \ref{app: retarded} for the detail).
We can analytically continue the expression with respect to $\Delta$, and we obtain a finite result even for $\Delta>1$, which agrees with $\Pi$ in \eqref{eq: final Pi}.
The analytically continued one is no longer the same as $\braket{O}_t$ in CFT, which as explained above generally diverges for $\Delta >1$ and depends on the regulator. 
This is the reason why the second law is violated in bulk for $1<\Delta <2$ as figure~\ref{fig: bulk second}, although it holds in CFT.
Thus, the holographic renormalization scheme that we used is unsuitable for the second law which must hold in well-defined CFTs because it is a consequence of the non-negativity of the relative entropy. 
In addition, this consequence is not a problem of the invalidity of the holographic dictionary but an issue of the regularization, because as mentioned above we only see the universal parts.

The holographic renormalization can also be understood as dimensional regularization as follows.\footnote{This fact is already noted in \cite{Berenstein:2014cia} as ``the AdS gravity computation already knows about
dimensional regularization''.} 
In general dimension $d$ for CFT, $\braket{O}_t$ is finite for $0<\Delta <d/2$. 
We can then obtain the finite result for $\Delta> d/2$, by analytically continuing $\Delta$ or $d$. 
Thus, we can regard the above finite result in $\Delta>1$ as the analytical continuation of $d$, i.e., dimensional regularization.
 
In short, using the counterterms as in \cite{deHaro:2000vlm} naturally led us to adopt the dimensional regularization.
Then, the second law is violated for $\Delta>1$ with $\Pi(t)$ in \eqref{eq: final Pi}, and we are to conclude that the conventional holographic renormalization scheme is incompatible with the fundamental property of quantum mechanics, i.e., the non-negativity of relative entropy.

We further explain the equivalence of $\Pi(t)$ and $\braket{O}_t$ in appendix \ref{app: retarded}, where we show that they are the same only after the source is turned off, regardless of $\Delta$.

\subsection{Two coupled systems}\label{subsec: double example}
We consider two copies of \eqref{eq: example action} with different constants and couple them as section \ref{subsec: taking dual}.
We distinguish the two theories by using the superscript $\ith{1}$ and $\ith{2}$ where we regard $\ith{1}$ as the target system and $\ith{2}$ as the bath.
The source $w$ is turned off in the second theory.
This time, the boundary conditions for the fields $\Phi^\ith{i}$ are not the conventional Dirichlet ones as \eqref{eq: Phi BC}, but given as \eqref{eq: boundary conditions}, i.e.,
\begin{align}\label{eq: Phi12 BC}
  \Phi^\ith{1} \underset{r\to \infty}{\sim} - \left(\frac{(L^\ith{1})^2}{r} \right)^{2-\Delta^\ith{1}}\left(w + v \Pi^\ith{2} \right),\qquad
  \Phi^\ith{2} \underset{r\to \infty}{\sim} - \left(\frac{(L^\ith{2})^2}{r} \right)^{2-\Delta^\ith{2}}v \Pi^\ith{1}.
\end{align}
As explained in section~\ref{subsec: taking dual}, the two theories interact with each other only through these boundary conditions.
We again treat only the $s$-wave by taking the source $w(t)$ uniform in the spatial direction, and set the coupling $v$ constant for simplicity.

We now have two independent perturbation parameters $w(t)$ and $v$, and thus we have a variety of perturbative expansions. 
A simple choice is assuming that both of them are in the same order as $w=\mathcal{O}(\lambda), v=\mathcal{O}(\lambda)$. 
Then we can repeat a similar perturbative construction of the solution as in section~\ref{subsec: single example}, starting with the $\cO(\lambda^0)$ BTZ geometry for both of the two theories.
However, this choice is not interesting, because taking $C^\ith{1,2} = \cO(\lambda^0)$ as in the previous section, it will turn out that\footnote{As in the previous example, we regard $\kappa$ and $\lambda$ are independent parameters.} 
\begin{align}
  \Phi^\ith{1} = \cO(\lambda),\qquad
  \Phi^\ith{2} = \cO(\lambda^2),\qquad
  \Pi^\ith{1} = \cO(\lambda),\qquad
  \Pi^\ith{2} = \cO(\lambda^2).
\end{align}
Thus, $\Phi^\ith{2}$ is suppressed compared to $\Phi^\ith{1}$, and the leading correction to the entropy comes only from the excitation of $\Phi^\ith{1}$. 
Then, the result is completely the same as the previous one if we consider up to $\mathcal{O}(\lambda^2)$ because the entropy production for the bath system is $\mathcal{O}(\lambda^3)$.

The result changes if we consider a case that two terms $w$ and $v \Pi^\ith{2}$ for the asymptotic values of $\Phi^\ith{1}$ in \eqref{eq: Phi12 BC} are in the same order. 
This situation can be realized by taking, for example, 
\begin{align}
  w=\mathcal{O}(\lambda), \quad v=\mathcal{O}(\lambda^s),\quad
  C^\ith{1} = \cO(\lambda^0),\quad
  C^\ith{2} = \cO(\lambda^{-2s}),
\end{align}
where $s$ is an arbitrary real number (and indeed $s$ can be zero which may be the most natural because the coupling $v$ does not have to be correlated with the strength $\lambda$ of the protocol $w$ which is tunable by hand).
In this choice, from \eqref{eq: example Pi}, we find
\begin{align}
  \Phi^\ith{1} = \cO(\lambda),\qquad
  \Phi^\ith{2} = \cO(\lambda^{1+s}),\qquad
  \Pi^\ith{1} = \cO(\lambda),\qquad
  \Pi^\ith{2} = \cO(\lambda^{1-s}).
\end{align}
This indicates that $C^\ith{2}$ times the bulk energy momentum tensor of the bath system is $\mathcal{O}(\lambda^2)$, and hence the same perturbative construction of the solution as in section \ref{subsec: single example} is valid as well for the bath.

Performing the perturbative computation to $\mathcal{O}(\lambda^2)$, we find that the entropy density of the system and of the bath in this case are
\begin{align}
  \frac{s^\ith{1}(t)}{2\pi} &= \frac{ r_+^\ith{1}}{\kappa^\ith{1}} - \frac{(L^\ith{1})^3}{r_+^\ith{1}} \int_0^t \d t' \left(w + v \Pi^\ith{2} \right)\dot \Pi^\ith{1},\label{eq: s1}
  \\
  \frac{s^\ith{2}(t)}{2\pi} &= \frac{r_+^\ith{2}}{\kappa^\ith{2}} - v\frac{(L^\ith{2})^3}{r_+^\ith{2}} \int_0^t \d t'\, \Pi^\ith{1}\dot \Pi^\ith{2},\label{eq: s2}
\end{align}
where $\Pi^\ith{i}$ is computed as
\begin{align}
  &\Pi^\ith{1}(t) = \int\d \omega e^{-i\omega t}\Pi_\omega^\ith{1},\qquad
  \Pi^\ith{2}(t) = v \int\d \omega e^{-i\omega t} \Pi_\omega^\ith{1} J^\ith{2}_\omega,\\
  &J^\ith{i}_\omega := \frac{2 C^\ith{i}L^\ith{i}\Gamma (2-\Delta^\ith{i})^2}{\pi \kappa^\ith{i} }\left(\frac{r_+^\ith{i}}{(L^\ith{i})^2} \right)^{2\Delta^\ith{i}-2}
  (-\sin(\pi \Delta^\ith{i}))
  \left(\frac{\Gamma(\alpha_\omega^\ith{i} )}{\Gamma(\gamma_\omega^\ith{i} - \alpha_\omega^\ith{i})} \right)^2,\\
  &\Pi_\omega^\ith{1} := \frac{w_\omega J^\ith{1}_\omega}{1 - v^2 J^\ith{1}_\omega J^\ith{2}_\omega}.
\end{align}

The second law, $s^\ith{1}(t) + s^\ith{2}(t) \geq s^\ith{1}(0)+s^\ith{2}(0)$, claims that,
\begin{align}\label{eq: example total second law}
  0\le - \frac{(L^\ith{1})^3}{r_+^\ith{1}} \int_0^t \d t'\, \left(w + v \Pi^\ith{2} \right)\dot \Pi^\ith{1} - \frac{(L^\ith{2})^3}{r_+^\ith{2}} \int_0^t \d t'\, (v\Pi^\ith{1})\dot \Pi^\ith{2}.
\end{align}
Below, we are going to show that the above inequality holds as far as each theory satisfies \eqref{eq: single second law}.
Let us define $f(t)$ by
\begin{align}
  f(t) := \int \d \omega\, e^{-i \omega t}\, \left(w_\omega + v^2 \Pi_\omega^\ith{1} J^\ith{2}_\omega \right)
 = \int \d \omega \frac{e^{-i \omega t}w_\omega}{1 - v^2 J_\omega^\ith{1} J_\omega^\ith{2}},
\end{align}
and write its Fourier coefficient as $f_\omega$.
Then, we obtain
\begin{align}\label{eq: omega by f}
  w(t) = f(t) - v^2 \int\d \omega\, e^{-i\omega t} J_\omega^\ith{1}J_\omega^\ith{2}f_\omega.
\end{align}
Instead of giving $f$ by $w$, we here rather define the source $w$ by this expression, choosing $f$ as any Schwartz function.
As far as the support of $f$ is set to be $\mathbb{R}^+$, that of $w$ is guaranteed to be so too, since there is no pole on the upper half $\omega$ plane in the integral of \eqref{eq: omega by f}.
Then, from the relations,
\begin{align}
  w(t) + v \Pi^\ith{2}(t) = f(t),\qquad
  \Pi_\omega^\ith{1} = f_\omega J_\omega^\ith{1},
\end{align}
the inequality \eqref{eq: single second law} can be applied to the first term in \eqref{eq: example total second law} with $w$ replaced by $f$.
We can also show the same thing for the second term in \eqref{eq: example total second law} with $w$ replaced by $v\Pi^\ith{1}$.
Since each term in the r.h.s.\ of \eqref{eq: example total second law} is not negative if each theory satisfies \eqref{eq: single second law} for general sources $w(t)$, \eqref{eq: example total second law} itself is now guaranteed.

\section{Summary and discussions}\label{sec: summary}
We have introduced a system consisting of AdS black holes interacting via the conformal boundary, which was shown to be dual to coupled holographic CFTs. 
The properties in the black hole thermodynamics were first derived in the CFT language, and after that, translated to the bulk based on the AdS/CFT dictionary \cite{Gubser:1998bc, Witten:1998qj}.
Here, the coarse-grained state on the boundary corresponds to auxiliary Euclidean black holes that share the same asymptotic observables with the original Lorentzian spacetimes.
The second law states that the sum of the horizon areas of the Euclidean black holes that coarse-grain the system at time $t$ can never get smaller than its value at $t=0$.
When one is seen as the target system and the others as the baths, the notions of heat and work naturally come in the first and the second laws, making it possible to discuss the black hole thermodynamics more in parallel with the ordinary thermodynamics than before.

Also, we have explicitly checked whether the second law is actually satisfied in a gravity model. 
We have demonstrated that the second law is satisfied for the model in a specific parameter region. 
However, we have encountered a violation of the second law in another parameter region. 
We have argued that the violation is related to a problem of UV divergences.
In this parameter region, responses in the CFT generally have UV divergences.
Since the bulk counterparts are calculated to be finite due to the holographic renormalization scheme, it has turned out that the renormalization scheme that we used breaks the second law.
In our example, we adopted the counterterms proposed in \cite{deHaro:2000vlm}, and argued that it corresponds to dimensional regularization.
Surprisingly, dimensional regularization is incompatible with the fundamental property of quantum mechanics: non-negativity of the relative entropy. 
It is an important subject to discover a renormalization scheme that respects this property.

In our strategy, there was no need to refer to energy conditions to prove the second law since we instead used the holographic dictionary unlike other strategies.
As the boundary theory is expected to know the bulk UV in principle, our second law would contain some information about the quantum gravity, which might have allowed us to derive the second law without assuming energy conditions (after we resolve the puzzle about the renormalization discussed above).
Then, how can we extract messages on quantum gravity from this second law?
One possible idea is to use it as a judge of whether or not the gravity is UV-complete.
We can check the second law for models like the way we did in section \ref{sec: example}, using it as a necessary condition.
However, it is not yet guaranteed that all UV complete gravity theories share the same second law, with or without holography, so we need to study more on this point.

We can consider various extensions of the examples in section~\ref{sec: example}. 
For simplicity, we have considered only spatially uniform sources in two-dimensional CFT for the examples. 
Due to this, the entropy production does not depend on the details of the theories because the production is, at the leading order of the perturbation, determined by the two-point function which is fixed by the conformal symmetry. 
To see more interesting cases, we should consider non-uniform sources or higher-dimensional models. 
In addition, in the concrete examples, we have chosen only the Hamiltonians as the operators to be respected, although we have investigated general prescriptions in section~\ref{sec: bulk}.
One of the simple extensions is respecting also conserved charges such as electric charges and angular momenta. 
In two-dimensional CFTs, we can also respect the quantum KdV charges generically \cite{Maloney:2018hdg}. Then, the coarse-grained states are given by generalized Gibbs ensembles and they can be dual to non-trivial black holes (KdV-charged black holes) \cite{Dymarsky:2020tjh}.
It is interesting to study the entropy production for these charged black holes. 
Furthermore, in our approach, we can respect operators not commuting with Hamiltonians. 
It is an important open problem to find non-trivial black hole backgrounds corresponding to this case.

In the example in section \ref{subsec: double example}, we have confirmed that the second law of the total entropy is trivially satisfied in the sense that the entropy of each theory individually increases.
One of the most interesting examples is the case where the entropy of a target system decreases owing to the introduction of the bath.
Since we have only examined for constant $v$, there may still be room for the theory in section \ref{sec: example} to be such an example for some varying $v$.
Extending the analysis beyond the perturbative computations may also provide examples where the entropy of the target system decreases.
If such an example is found, we will get closer to the realization of a ``black hole engine".
Operating the black hole engine governed by general relativity, we will be able to describe the dual quantum thermodynamic engine in laboratories with the philosophy of \cite{Hashimoto:2022aso, Hashimoto:2024yev}.

The black hole engine would not only be beneficial in understanding the boundary theory, but also the bulk gravity.
If generalized to the information thermodynamics, the engine will help us understand the connection between spacetime geometry and the notion of information that Maxwell's demon acquires in the measurement.
After the discovery of the Ryu-Takayanagi formula \cite{Ryu:2006bv, Ryu:2006ef}, quantum information became one of the insightful tools to study quantum gravity.
Here, we expect that the information thermodynamic engine provides other aspects of the relation between geometry and information.

Finally, there is a technical point that we need to survey more.
As pointed out in the erratum of \cite{Takeda:2024qbq}, there is an ambiguity about the dictionary \eqref{eq: EM expectation value extracted}.
When $\mathrm{U}(1)$ current operator that couples to the external gauge field is considered, the bulk theory contains the Maxwell field.
Then, particularly for the charged static black hole, the source is associated with the chemical potential, but this time, it is rather consistent if we ignore the second term of $\braket{T^\ith{s}_{\mu\nu}}$ in \eqref{eq: EM expectation value extracted}, implying that $Y_{\mu\nu }$ does not include the chemical potential term from the beginning.
This also seems to happen for the time component of other tensor fields.
It must be more preferable if there exists a dictionary that applies to all CFT operators and corresponding bulk fields.

\subsection*{Acknowledgment}
We thank Koji Hashimoto, Takanori Ishii, Norihiro Tanahashi, and Ryota Watanabe for the discussions.
D.T.\ appreciates Yu Nakayama, Yuki Suzuki and Seiji Terashima for the comments and discussions in the seminar at YITP.
The work of T.S.\ was supported by JST SPRING, Grant Number JPMJSP2110.
S.S.\ acknowledges support from JSPS KAKENHI Grant Numbers JP21K13927 and JP22H05115.
The work of D.T.\ is supported by Grant-in-Aid for JSPS Fellows No.\ 22KJ1944.
The work of T.Y.\ was supported in part by JSPS KAKENHI Grant No.\ JP22H05115 and JP22KJ1896.

\appendix

\section{Generalization to multiple interacting theories}\label{app: multiple version}
In this appendix, we generalize section \ref{sec: boundary} and \ref{sec: bulk} to the multiple interacting theories.
In addition, we consider more operators to be respected.

\subsection{Generalization of section \ref{sec: boundary}}\label{subapp: sec 2}
We consider a generic setup of $M$ quantum systems coupled to each other:
\begin{align}
  H_\ttl(t) = \sum_{i=1}^M H^\ith{i}(w^\ith{i}(t)) + \sum_{i<j} V^\ith{ij}(t),
\end{align}
where $H^\ith{i}$ is the time-dependent Hamiltonian of the $i$-th system whose protocol is specified by $w^\ith{i}(t)$,
and where $V^\ith{ij}(t)$ is the interaction between the $i$-th and the $j$-th systems.
For the vanishing protocol $w^\ith{i}(t) \equiv 0$, the Hamiltonian $H^\ith{i}(w^\ith{i})$ becomes time-independent, $H^\ith{i}(0)=:H^\ith{i}_*$.
The initial state is set to be $\rho(0)$, which evolves as
\begin{align}
  \rho(t) = U(t) \rho(0) U(t)^\dagger,\qquad
  U(t) := \mathrm{T}\exp\l(-i\int_0^t\d t'\, H_\ttl(t')\r).
\end{align}

In the coarse-graining, let $\{O_J^\ith{i}\}$ be the set of operators that we respect in each $i$-th theory, where $J$ is the label of the operators in this set.
The coarse-grained state can be found as we did in section \ref{sec: boundary}:
\begin{align}
    \bar \rho(t) = \bigotimes_{i=1}^M \bar\rho^\ith{i}(t),\qquad
    \bar \rho^\ith{i}(t) =\frac{1}{Z^\ith{i}(t)} \exp\l(-\sum_J \lambda_{J}^\ith{i}(t)O_J^\ith{i}\r),
\end{align}
with $\Tr_i$ denoting the trace over the $i$-th degrees of freedom.
The Lagrange multipliers $\lambda^\ith{i}_{J}(t)$ are determined so that $\bar\rho^\ith{i}(t)$ is the maximum entropy state with respecting the expectation values $\braket{O^\ith{i}_J}_t$ (see \eqref{eq: QM expectation value} and \eqref{eq: coarse-grained conditions}).
We define the $i$-th coarse-grained entropy $S^\ith{i}(t)$ at time $t$ as
\begin{align}
  S^\ith{i}(t) := -\Tr_i\l[\bar\rho^\ith{i}(t) \ln \bar\rho^\ith{i}(t)\r] = \sum_J\lambda_{J}^\ith{i}(t)\braket{O_J^\ith{i}}_t + \ln Z^\ith{i}(t).
\end{align}
The total entropy is defined as the sum of them,
\begin{align}
  S(t) = \sum_i S^\ith{i}(t).
\end{align}

We obtain the second law and the fundamental relation, similarly as
\begin{align}
    &S(t) \geq S(0),\\
    &\dot S^\ith{i} (t)= \sum_J \lambda_{J}^\ith{i}(t) \frac{d }{d t} \braket{O_J^\ith{i}}_t,
\end{align}
imposing again $\rho(0) = \bar \rho(0)$.
In a more specific case where $O_{J=0}^\ith{1}=H_*^\ith{1}$ and only the Hamiltonian is respected for each $i$-th ($i\geq2$) system, i.e., $\{O^\ith{i}_{J}\} = \{H_*^\ith{i}\}$, the above, as \eqref{eq: 2nd law with heat}, reduces to
\begin{align}
  &S^\ith{1}(t) \geq S^\ith{1}(0) + \sum_{i=2}^M \int_0^t\d t'\, \beta^\ith{i}(t')\delta Q^\ith{i}(t'),
  \qquad
  \delta Q^\ith{i}(t) := - \frac{d }{d t} \braket{H_*^\ith{i}}_t,\label{eq: special multiple second}
  \\
  & \dot S^\ith{1}(t) = \beta^\ith{1}(t)\l[
  	\frac{d }{d t}\braket{H_*^\ith{1}}_t - \sum_{J \neq 0} \mu_{J}^\ith{1}(t)\, \frac{d}{d t} \braket{O_J^\ith{1}}_t
  \r].\label{eq: special multiple fundamental relation}
\end{align}
Here, we have put $\lambda_{J=0}^\ith{i}(t) = \beta^\ith{i}(t)$ and $\lambda_{J}^\ith{1}(t) = -\beta^\ith{1}(t) \mu_{J}^\ith{1}(t)$ for $J\neq 0$.

\subsection{Generalization of section \ref{sec: bulk}}\label{subapp: taking dual}
We consider the gravitational model corresponding to the above specific situation, i.e., only the Hamiltonian is respected for theories of $i\geq 2$ except for the first theory ($i=1$).
The action \eqref{eq: CFT action} can be generalized as
\begin{align}\label{appeq: CFT action}
  I_\ttl = \sum_{i=1}^M I^\ith{i}[\gamma^\ith{i}] - \sum_J \int\d^d x  \sqrt{-\gamma^\ith{1}}\, O^\ith{1}_J(x) \left[w_J(x)+ \sum_{j = 2}^M v_J^\ith{j}(x) O^\ith{j}_1(x) \right],
\end{align}
where $i, j$ distinguishes the theories and $J$ is the operator label for the 1st theory.
We denote the operator in $j$-th theory $O^\ith{j}_1(x)$ rather than $O^\ith{j}(x)$ for later convenience.
For simplicity, we suppose that theories of $j\ge 2$ do not interact directly with each other.
See appendix \ref{app: three CFT} for the case of three mutually interacting theories.
Below, we use the dot product to represent the sum over $J$, like $w\cdot O^\ith{1} = \sum_J w_J O^\ith{1}_J$.

\subsubsection*{Holographic dictionaries}
We perform the path integral \eqref{eq: CFT Z} for the above action with $\mathrm{D}\varphi := \mathrm{D}\varphi^\ith{1}\cdots\mathrm{D}\varphi^\ith{M}$.
By introducing auxiliary fields following section \ref{subsec: taking dual}, we obtain
\begin{align}
    &Z[\gamma, w, v]\nonumber\\
    &=\int \mathrm{D}\varphi \mathrm{D}\eta \mathrm{D}\chi \exp\left[i \sum_{i=1}^M I^\ith{i} + i \int\d^d x  \sqrt{- \gamma^\ith{1}}\left\{\eta \cdot  \left(v^\ith{j}O^\ith{j}_1- \chi \right) -O^\ith{1}\cdot  \left(w^\ith{1}+ \chi \right) \right\} \right]\nonumber\\
    &=\int \mathbf{D}\Phi \mathrm{D}\eta \mathrm{D}\chi
  \exp\left[i\sum_{i=1}^M\mathcal{I}^\ith{i} - i \int\d^d x  \sqrt{- \gamma^\ith{1}}\,\eta \cdot \chi \right]
  \prod_{J} \delta\left(\hat \Phi^\ith{1}_J + w_J^\ith{1} + \chi_J \right)\nonumber\\
  &\hspace{120pt}
  \times
  \prod_{j=2}^M \left[\delta\left(\hat \Phi^\ith{j}_1 - \eta \cdot  v^\ith{j} \right)\prod_{K \neq 1}\delta \left(\hat \Phi^\ith{j}_K \right) \right]
  \prod_{k=1}^M\delta \left(\hat g^\ith{i}_{\mu\nu } - \gamma_{\mu\nu }^\ith{i} \right),\nonumber\\
  &=\int \mathbf{D}\Phi \mathrm{D}\eta
  \exp\left[i\sum_{i=1}^M\mathcal{I}^\ith{i} + i \int\d^d x  \sqrt{- \gamma^\ith{1}}\,\eta \cdot \left(\hat \Phi^\ith{1} + w^\ith{1} \right) \right]\nonumber\\
  &\hspace{120pt}
  \times
  \prod_{j=2}^M \left[\delta\left(\hat \Phi^\ith{j}_1 - \eta \cdot  v^\ith{j} \right)\prod_{K \neq 1}\delta \left(\hat \Phi^\ith{j}_K \right) \right]
  \prod_{k=1}^M\delta \left(\hat g^\ith{i}_{\mu\nu } - \gamma_{\mu\nu }^\ith{i} \right).
\end{align}
In the above, the bulk metrics are as before written as
\begin{align}
    G^\ith{i}_{MN}\d X^M \d X^N = \left(\frac{L^\ith{i}}{z} \right)^2\d z^2 + g^\ith{i}_{\mu\nu }\d x^\mu \d x^\nu,
\end{align}
and the hat quantities are defined by
\begin{align}
  g^\ith{i}_{\mu\nu } \sim \left(\frac{L^\ith{i}}{z} \right)^2\hat g^\ith{i}_{\mu\nu },\qquad
  \Phi_J^\ith{i} \sim z^{d-\Delta_J^\ith{i}}\hat \Phi_J^\ith{i}\qquad (z\to 0).
\end{align}
Here, $\Delta_J^\ith{i}$ is the conformal dimension of boundary operator $O_J^\ith{i}$.
In addition, $\Phi^\ith{j}_K$'s ($K\neq 1$) denote the other bulk fields whose dual operators do not appear in the interaction term of \eqref{appeq: CFT action}.

Taking the large $N$ limit, we are left with the classical action
\begin{align}
  \mathcal{I}_\ttl := \sum_{i=1}^M\mathcal{I}^\ith{i} + \int\d^d x  \sqrt{- \gamma^\ith{1}}\,\eta \cdot \left(\hat \Phi^\ith{1} + w^\ith{1} \right),
\end{align}
with the boundary conditions,
\begin{align}
    0 = \hat \Phi^\ith{j}_1 - \eta \cdot  v^\ith{j} = \hat \Phi^\ith{j}_{K\neq 1} = \hat g^\ith{i}_{\mu\nu } - \gamma_{\mu\nu }^\ith{i}
  \qquad
  (i = 1,\cdots, M,~ j = 2,\cdots,M).
\end{align}
Then, the variations on the boundary are restricted as
\begin{align}
  0 = \delta \hat \Phi^\ith{j}_1 -\delta \eta \cdot  v^\ith{j} = \delta \hat \Phi^\ith{j}_{K \neq 1} = \delta \hat g^\ith{i}_{\mu\nu }.
\end{align}
Noting those conditions, we require
\begin{align}
 0= \delta \mathcal{I}_\ttl &= (\mbox{EOMs}) + \int\d^d x\sqrt{-\gamma^\ith{1}}\left[\left(\Pi^\ith{1} + \eta \right)\cdot \delta \hat\Phi^\ith{1} + \left( \Pi_1^\ith{j} v^\ith{j} + \hat \Phi^\ith{1} + w^\ith{1} \right)\cdot \delta \eta \right],\nonumber\\
 \mathrm{i.e.,}\quad
 0&=\Pi^\ith{1}_J + \eta_J = \Pi_1^\ith{j} v^\ith{j}_J + \hat \Phi^\ith{1}_J + w^\ith{1}_J,
\end{align}
where we have defined
\begin{align}
  \Pi_J^\ith{i}(x) : =  \frac{1}{\sqrt{-\gamma^\ith{i}}}\frac{\delta \mathcal{I}^\ith{i}|_\mathrm{on-shell}}{\delta \hat \Phi_J^\ith{i}(x)}.
\end{align}
Therefore, by eliminating $\eta$, the remaining task is to find the solutions subject to
\begin{align}\label{appeq: bulk BC}
  0 = \hat \Phi_J^\ith{1} + w_J^\ith{1} + \sum_{k=2}^M v_J^\ith{k} \Pi_1^\ith{k} =  \hat \Phi^\ith{j}_1 + v^\ith{j} \cdot \Pi^\ith{1}  = \hat \Phi^\ith{j}_{K\neq 1} = \hat g^\ith{i}_{\mu\nu } - \gamma_{\mu\nu }^\ith{i},
\end{align}
and to evaluate the generating functional with the on-shell bulk action,
\begin{align}
  -i \ln Z[\gamma,w,v] = \mathcal{I}_\ttl|_\mathrm{on-shell} = \sum_{i=1}^M\mathcal{I}^\ith{i} + \sum_{k=2}^M \int\d^d x  \sqrt{- \gamma^\ith{1}}\,\left(v^\ith{k} \cdot \Pi^\ith{1} \right) \Pi_1^\ith{k}.
\end{align}

The dictionaries on the expectation values can be derived in the same way as section \ref{subsec: taking dual}:
\begin{align}
    \braket{O_J^\ith{1}(x)} = \Pi_J^\ith{1}(x),\qquad
    \braket{T_{\mu\nu }^\ith{i}} = Y_{\mu\nu }^\ith{i} + \delta^{1i} w^\ith{1}\cdot \Pi^\ith{1} \gamma_{\mu\nu }^\ith{1},
\end{align}
with the Brown-York tensor defined as
\begin{align}
    Y^\ith{i}_{\mu\nu }(x) := -\frac{2}{\sqrt{-\gamma^\ith{i}}} \frac{\delta \mathcal{I}^\ith{i}}{\delta \gamma_\ith{i}^{\mu\nu }(x)}.
\end{align}
We hereafter set $\gamma^\ith{i} = \gamma$, which was defined in \eqref{eq: boundary metric}.

\subsubsection*{Building the bulk}
We construct the dual bulk time evolution.
The operators that we respect here are $H^\ith{i}_*$, $O_J^\ith{1}$, and additionally, the momentum operator
\begin{align}
    P_a^\ith{1} = \int d^{d-1}\vec x \sqrt{\sigma} T^\ith{1}\.^0_{\phantom{0}a},
\end{align}
where $\sigma$ is the spatial metric on the boundary given by \eqref{eq: boundary metric}.
This operator enables us to respect the angular momentum in the dual bulk.
Since there is no change on the bath theories $j \ge 2$, we here focus on the first theory.

At $t=0$, we prepare the initial state as a generalized Gibbs state,
\begin{align}
    \rho^\ith{1}(t) &= \frac{1}{Z^\ith{1}(t)} \exp\left[-\beta(0) \left(H^\ith{1}_* - \omega^a(0) P_a^\ith{1} - \int \d^{d-1} \vec x\sqrt{\sigma} \mu(0,\vec x)\cdot  O^\ith{1}(\vec x) \right) \right],\\
    Z^\ith{1}(t) &= \Tr_1\exp\left[-\beta(0) \left(H^\ith{1}_* - \omega^a(0) P_a^\ith{1} - \int \d^{d-1} \vec x\sqrt{\sigma} \mu(0,\vec x)\cdot  O^\ith{1}(\vec x) \right) \right].
\end{align}
Here we have put $\beta^\ith{1} = \beta$ for notational simplicity.
As explained in section \ref{subsec: BHT}, the Hamiltonian of $t<0$ must be chosen as
\begin{align}
    H_\ttl = \sum_{i=1}^M H_*^\ith{i} - \omega^{a}(0) P_a^\ith{1} - \int \d^{d-1} \vec x\sqrt{\sigma} \mu^\ith{1}(0,\vec x)\cdot  O^\ith{1}(\vec x)\qquad (t<0).
\end{align}
The main difference between this and \eqref{eq: t<0 Hamiltonian} is the addition of the momentum operator.
The combination $H_*^\ith{1} - \omega^a(0)P_a^\ith{1}$ can be regarded as the Hamiltonian when the metric is given as
\begin{align}\label{eq: twisted past boundary metric}
 \d s^2_\ith{1} =  -\d t^2 + \sigma_{ab}\left(\d x^a + \omega^{a}(0)\d t \right)\left(\d x^b + \omega^{b}(0)\d t \right)\qquad
 (t<0).
\end{align}
This is because we see
\begin{align}
  H^\ith{1}_* - \omega^{a}(0) P_a^\ith{1} = -\int \d^{d-1}\vec x\sqrt{\sigma}\,  T^\ith{1}\.^0_{\phantom{0}\mu} \left(\delta^\mu_0 + \omega^{a}(0) \delta^\mu_a \right),
\end{align}
and it is, from the ADM formalism, nothing but the Hamiltonian with the shift vector chosen as $\omega^{a}(0)$.
Since the metric is just a source and not dynamical, we need not worry about any geometric singularity due to the change of the metric at $t=0$.
Note that the determinant of \eqref{eq: twisted past boundary metric} is the same as $\sqrt{-\gamma}$.

Therefore, the bulk of the first theory is constructed as follows.
For $t<0$, we find the stationary solution that matches the asymptotic boundary conditions,
\begin{align}
    \hat \Phi_J^\ith{1}(t,\vec x) = \mu_{J}(0,\vec x),\quad
  \hat g^\ith{1}_{\mu \nu }\d x^\mu \d x^\nu  = -\d t^2 + \sigma_{ab}\left(\d x^a + \omega^{a}(0)\d t\right)\left(\d x^b + \omega^{b}(0)\d t \right).
\end{align}
Then, we solve the initial value problem from $t=0$ toward future, with the asymptotic boundary conditions \eqref{appeq: bulk BC} and \eqref{eq: boundary metric}.

\subsubsection*{Coarse-graining the dynamical black holes}
We continue to focus on the first theory.
The coarse-grained state in the CFT side is this time given as
\begin{align}
    \bar\rho^\ith{1}(t) &= \frac{1}{Z^\ith{1}(t)} \exp\left[-\beta(t) \left(H^\ith{1}_* - \omega^{a}(t) P_a^\ith{1} - \int \d^{d-1} \vec x\sqrt{\sigma} \mu(t,\vec x)\cdot  O^\ith{1}(\vec x) \right) \right],\\
    Z^\ith{1}(t) &= \Tr_1\exp\left[-\beta(t) \left(H^\ith{1}_* - \omega^{a}(t) P_a^\ith{1} - \int \d^{d-1} \vec x\sqrt{\sigma} \mu(t,\vec x)\cdot  O^\ith{1}(\vec x) \right) \right].
\end{align}
The bulk description of $Z^\ith{1}(t)$ is a stationary Euclidean solution that matches the following asymptotic boundary conditions:
\begin{align}
  \hat \Phi_J^\ith{1}(\tau, \vec x) = -\mu_{J}(t,\vec x),\quad
  \hat g^\ith{1}_{\mu \nu }\d x^\mu \d x^\nu  = \d \tau^2 + \sigma_{ab}\left(\d x^a - i \omega^a(t)\d \tau \right)\left(\d x^b - i \omega^b(t)\d \tau \right).
\end{align}
Notice that the Lorentzian time $t$ and the Euclidean time $\tau$ are independent and distinguished.

\subsubsection*{The coarse-grained entropy}
For all $i$, the coarse-grained entropy can be rewritten as
\begin{align}
  S^\ith{i}(t) = -\left(\beta^\ith{i}(t) \right)^2\frac{ \partial}{\partial \beta^\ith{i}(t)}\left[\left(\beta^\ith{i}(t) \right)^{-1} \ln Z^\ith{i}(t) \right].
\end{align}
If $\mathcal{I}^\ith{i}$ is the Einstein theory, the bulk second law reads,
\begin{align}
     \sum_{i=1}^M \frac{A^\ith{i}(t)}{4G^\ith{i}} \geq \sum_{i=1}^M \frac{A^\ith{i}(0)}{4G^\ith{i}}.
\end{align}
Each $A^\ith{i}(t)$ is the horizon area of the $i$-th Euclidean black hole at $t$, and $G^\ith{i}$ is the Newton constant of $\mathcal{I}^\ith{i}$.
Alternatively, we also obtain $\varsigma(t) \ge 0$ for
\begin{align}
    \varsigma(t) &:= \frac{A^\ith{1}(t)}{4G^\ith{1}} - \frac{A^\ith{1}(0)}{4G^\ith{1}} + \sum_{j=2}^M \int_0^t\d t'\,\beta^\ith{j}(t')\dot M^\ith{j}(t'),\\
  M^\ith{j}(t) &:= \int \d^{d-1}\vec x \sqrt{\sigma}\, Y^\ith{j}_{00}.
\end{align}

\subsection{Holographic dual of three mutually coupled CFTs}\label{app: three CFT}
As another generalization of section \ref{subsec: taking dual} and appendix \ref{subapp: taking dual}, we consider the following action:
\begin{align}\label{eq: appendix Z}
  I_{\ttl} = \sum_{i=1}^3I^\ith{i} - \int\d ^dx\sqrt{-\gamma}\left[v_1 O^\ith{2} O^\ith{3}  + v_{2} O^\ith{3}O^\ith{1}+v_3 O^\ith{1} O^\ith{2}\right].
\end{align}
Using the same strategy and the notations in section \ref{subapp: taking dual}, we are going to move from the above path integral to its gravitational dual description.
For notational simplicity, we write as if there were only the three bulk fields which correspond to the three primary operators $O^\ith{i}$, but the other fields including the metrics are of course implicitly included as we carefully did in section~\ref{subapp: taking dual}.
Also, we here turn off the source terms.

The path integral \eqref{eq: CFT Z} with \eqref{eq: appendix Z} is rewritten as follows:
\begin{align}
&Z[v]\nonumber\\
 &= \int \mathrm{D}\varphi \mathrm{D}\eta \mathrm{D}\chi\,e^{i\sum_{i=1}^3I^\ith{i} }\nonumber\\ &\hspace{60pt}\times\exp\left[- i\int\d^d x\sqrt{-\gamma}\left(\sum_{j=2,3} \eta_j (\chi_j - O^\ith{j}) +  v_1 \chi_2 \chi_3 + (v_2 \chi_3 + v_3\chi_2) O^\ith{1} \right) \right]\nonumber\\
 &=\int \mathbf{D}\Phi \mathrm{D}\eta \mathrm{D}\chi \exp\left[i\sum_{i=1}^3\mathcal{I}^\ith{i} - i\int\d^d x\sqrt{-\gamma}\left(\eta_2\chi_2 + \eta_3\chi_3 +  v_1\chi_2 \chi_3 \right) \right]\nonumber\\
 &\hspace{180pt} \times \delta\left(\hat \Phi^\ith{1} + v_2\chi_3 +v_3 \chi_2 \right) \delta\left(\hat \Phi^\ith{2} -\eta_2 \right)\delta\left(\hat \Phi^\ith{3} - \eta_3 \right)\nonumber\\
 &=\int \mathbf{D}\Phi \mathrm{D}\chi \exp\left[i\sum_{i=1}^3\mathcal{I}^\ith{i} - i\int\d^d x\sqrt{-\gamma}\left(\chi_2\hat \Phi^\ith{2} + \chi_3 \hat\Phi^\ith{3} +  v_1 \chi_2\chi_3 \right) \right]\nonumber\\
 &\hspace{180pt} \times \delta\left(\hat \Phi^\ith{1} + v_2 \chi_3 + v_3 \chi_2 \right).
\end{align}
We cannot perform the $\chi$-integration in the final line, because $v_2$ and $v_3$ are not invertible ($v_2,v_3$ can be zero) in general.

Therefore, let us at this stage take the large $N$ limit.
Repeating what we have done in section \ref{subsec: taking dual}, we conclude that the bulk configuration is determined by solving the bulk EOMs with the following asymptotic boundary conditions:
\begin{align}
  0  = \hat \Phi^\ith{1} - v_2 \Pi^\ith{3} - v_3 \Pi^\ith{2}= \hat \Phi^\ith{2} - v_3 \Pi^\ith{1} - v_1 \Pi^\ith{3} = \hat \Phi^\ith{3} - v_1 \Pi^\ith{2} - v_2 \Pi^\ith{1}.
\end{align}
The auxiliary fields $\chi_2$ and $\chi_3$ have been eliminated through $\chi_2 + \Pi^\ith{2} = \chi_3 + \Pi^\ith{3}=0$.
The on-shell action is reduced to
\begin{align}
  \mathcal{I}_\ttl|_\mathrm{on-shell} = \sum_{i=1}^3 \mathcal{I}^\ith{i} + \int\d ^dx\sqrt{-\gamma}\left[v_1 \Pi^\ith{2} \Pi^\ith{3} + v_2 \Pi^\ith{3} \Pi^\ith{1} + v_3 \Pi^\ith{1} \Pi^\ith{2}\right].
\end{align}

The symmetry regarding the label $i=1,2,3$ is now recovered, although we broke it when introducing the auxiliary fields.
While we did not include the source terms, that is, the single-trace deformation terms, it is straightforward to take them into account.
Notice finally that since \eqref{eq: appendix Z} is a type of the theories that we have dealt with in section \ref{subapp: sec 2}, all the thermodynamic statements are also valid in this setup.
\section{QFT computations}\label{app: example in CFT}
Here we review a perturbative computation of the response under time-dependent perturbations, and apply it to two-dimensional CFTs.

\subsection{Perturbative computations}
We start at time $t_i$ with an initial density matrix $\rho(t_i)$ and consider the time evolution by a perturbed Hamiltonian 
\begin{align}
    H(t)=H_0+V(t),
\end{align}
where $H_0$ is the unperturbed Hamiltonian and $V(t)$ is the perturbative term.
Note that $H(t), V(t)$ are operators in the Schr\"odinger picture.
We then compute the expectation value of an operator $O$ at time $t_f$ ($t_f>t_i$):
\begin{align}\label{O(tf)}
    \braket{O}(t_f):=\tr(O\rho(t_f))=\tr(OU(t_f,t_i)\rho(t_i)U^{-1}(t_f,t_i)),
\end{align}
where $U$ is the time evolution operator defined as
\begin{align}
    U(t_1,t_2):=\mathrm{T}\exp\left(-i\int^{t_1}_{t_2}\!\!dt\, H(t)\right).
\end{align}
We also define the ``free'' time evolution operators as
\begin{align}
U_0(t_1,t_2):=e^{-iH_0(t_1-t_2)}. 
\end{align}

We now move to the interaction picture.
For any Schr\"odinger picture operator $A(t)$, the interaction picture operator $A^I(t)$ is defined  as
\begin{align}
    A^I(t):=U_0^{-1}(t,t_s)A(t)U_0(t,t_s),
\end{align}
where $t_s$ is an arbitrary reference time relating the two pictures.
The time evolution operator in the interaction picture is defined as 
\begin{align}
    U_I(t_1,t_2):=U_0^{-1}(t_1,t_s)U(t_1,t_2)U_0(t_2,t_s),
\end{align}
and can be computed as
\begin{align}
    U_I(t_1,t_2)&=T\exp\left(-i\int^{t_1}_{t_2}\!\!dt\, V^I(t)\right)
    \nn
    &=1-i\int^{t_1}_{t_2}\!\!dt\, V^I(t)-\int^{t_1}_{t_2}\!\!dt\,\int^{t}_{t_2}\!\!dt'\, V^I(t)V^I(t')+\mathcal{O}(V^3).\label{UIpert}
\end{align}

The expectation value \eqref{O(tf)} is computed, using 
\begin{align}
U(t_f,t_i)=U_0(t_f,t_s)U_I(t_f,t_i)U_0^{-1}(t_i,t_s)
\end{align}
and the perturbative expansion \eqref{UIpert},
as
\begin{align}
    \braket{O}(t_f)&=\tr(U_I^{-1}(t_f,t_i)O^I(t_f)U_I(t_f,t_i)U_0^{-1}(t_i,t_s)\rho(t_i)U_0(t_i,t_s))
    \nn
    &=\tr\left(O^I(t_f)\rho_0(t_i)\right)
    -i\int^{t_f}_{t_i}\!\!dt\tr\left([ O_I(t_f),V_I(t)]\rho_0(t_i)\right)
    \nn&\quad 
    -\int^{t_f}_{t_i}\!\!dt\,\int^{t}_{t_i}\!\!dt'\, \tr\left(
     [[O_I(t_f),V^I(t)],V^I(t')]\rho_0(t_i)\right)
     +\mathcal{O}(V^3),
     \label{Otf-pert}
\end{align}
where we have defined $\rho_0(t_i):=U_0^{-1}(t_i,t_s)\rho(t_i)U_0(t_i,t_s)$. 
The result \eqref{Otf-pert} can also be easily understood from the Schwinger-Keldysh path-integral.

The first term in \eqref{Otf-pert} is nothing but the expectation value of $O$ at $t_f$ for the unperturbed time evolution: 
\begin{align}
    \label{leadingO0}
    \braket{O}_0(t_f):=\tr\left(O^I(t_f)\rho_0(t_i)\right)=\tr\left(U_0^{-1}(t_f,t_i)OU_0(t_f,t_i)\rho(t_i)\right).
\end{align}
The second term in \eqref{Otf-pert} is the linear response of $O$ due to the perturbation $V$ (the Kubo formula):
\begin{align}\label{KubO}
    \delta \braket{O}(t_f):=-i\int^{t_f}_{t_i}\!\!dt\tr\left([ O_I(t_f),V_I(t)]\rho_0(t_i)\right).
\end{align}
Generically, this is the leading change of $\braket{O}$ under the perturbation $V$. 
However, if we set $O = H_0$ and the initial state $\rho(t_i)$ is static with respect to $H_0$, the linear response \eqref{KubO} vanishes. 
Indeed, this is the case because we will consider the energy change for the thermal initial state for $H_0$, and thus we have to consider the $\mathcal{O}(V^2)$ term in \eqref{Otf-pert}.

\subsection{Thermal initial state}
We consider, as the initial state $\rho(t_i)$, the thermal state of $H_0$ at inverse temperature $\beta$:
\begin{align}
    \rho(t_i)=\rho_\beta:=\frac{1}{Z(\beta)}e^{-\beta H_0}, \qquad Z(\beta):= \tr e^{-\beta H_0}. 
\end{align}
This state is static under the time evolution by $H_0$, and we have
\begin{align}
    \rho_0(t_i)=U_0^{-1}(t_i,t_s)\rho(t_i)U_0(t_i,t_s)=\rho_\beta.
\end{align}

When we add a perturbation, the time-evolved state $\rho(t)$ $(t>t_i)$ deviates from $\rho_\beta$, and we consider a coarse-grained state of $\rho(t)$ with respect to $H_0$:
\begin{align}
    \bar\rho(t)=\frac{1}{Z(\beta(t))}e^{-\beta(t) H_0},
\end{align}
where $\beta(t)$ is determined so that $\tr [H_0\bar\rho(t)]=\tr [H_0\rho(t)]$.

The leading change of the entropy $\delta S$ from $t_i$ to $t$ and that of the energy $\delta E$ are related as
\begin{align}
\label{S=betaE}
    \delta S= \beta \delta E,
\end{align}
where $\delta E=\tr [H_0\rho(t)]-\tr [H_0\rho(t_i)]$.
The second law states $\delta S \geq 0$ and thus we must have $\delta E \geq 0$.

Let us take the perturbation $V$ as
\begin{align}
    V(t)=w(t)\, O,
\end{align}
where $w(t)$ is the source for $O$.
Then, the leading change of $\braket{O}$ due to the perturbation is given by \eqref{KubO},
\begin{align}
    \delta \braket{O}(t_f)=-i\int^{t_f}_{t_i}\!\!dt\, w(t)\eqb{[ O_I(t_f),O_I(t)]}, \qquad \eqb{\bullet}:=\tr(\rho_\beta~ \bullet).
\end{align}
It can be written, if $w(t)=0$ for $t<t_i$, as
\begin{align}
    \delta \braket{O}(t_f)=\int^{\infty}_{-\infty}\!\!dt\,   w(t) G_R(t_f-t),
\end{align}
where $G_R$ is the thermal retarded Green's function
\begin{align}\label{eq: CFT retarded function}
    G_R(t):=-i \theta(t)\eqb{[ O_I(t),O_I(0)]}.
\end{align}
Note that the two-point function is time-translation invariant.
For later convenience, we also define a similar function (the Pauli-Jordan function or the invariant delta function) as
\begin{align}\label{eq: Pauli-Jordan}
    G(t):=-i \eqb{[ O_I(t),O_I(0)]}.
\end{align}
Using this, $\delta \braket{O}$ can also be written as
\begin{align}
    \delta \braket{O}(t_f)=\int^{t_f}_{-\infty}\!\!dt\,   w(t) G(t_f-t).
    \label{delOG}
\end{align}

The leading change of energy $\delta E$ is given by
\begin{align}
    \delta E(t_f)=
    -\int^{t_f}_{t_i}\!\!dt\,\int^{t}_{t_i}\!\!dt'\, w(t)w(t')\eqb{[[H_0, O_I(t)],O_I(t')]},
\end{align}
where the linear response term vanishes as stated above due to $[H_0, \rho_\beta]=0$.
Using the Heisenberg equation, $[H_0, O_I(t)]=-i \dot{O}_I(t)$, we obtain
\begin{align}
    \delta E(t_f)&=i\int^{t_f}_{t_i}\!\!dt\,\int^{t}_{t_i}\!\!dt'\, w(t)w(t')\eqb{[\dot{O}_I(t),O_I(t')]}
    \nn
    &=-\int^{t_f}_{-\infty}\!\!dt\,\int^{t}_{-\infty}\!\!dt'\, w(t)w(t')\frac{d}{dt}G(t-t')
    \nn
    &=-\int^{t_f}_{-\infty}\!\!dt\,w(t)\left(\frac{d}{dt}\int^{t}_{-\infty}\!\!dt'\, w(t')G(t-t')-w(t)G(0)\right)
    \nn
    &=-\int^{t_f}_{-\infty}\!\!dt\,w(t)\frac{d\delta \braket{O}}{dt}(t),
    \label{delEeq}
\end{align}
where we have used \eqref{delOG} and $G(0)=0$.
This agrees with the bulk calculation \eqref{eq: h_t} via \eqref{eq: O expectation value}.
Therefore, for any $t$ and any source $w(t)$, the second law claims
\begin{align}
    \int^{t}_{-\infty}\!\!dt'\,w(t')\frac{d\delta \braket{O}}{dt}(t') \leq 0,
\end{align}
which is of the same form as \eqref{eq: single second law}.
Recall that we have set $w(t)=0$ for $t<t_i$.

\subsection{Computations in two-dimensional CFT}
We will apply the previous result to two-dimensional CFTs. 
We consider a CFT with Hamiltonian $H_0$ and then add a perturbation as
\begin{align}
\label{CFT:V(t)}
    V(t)=\int dx\, w(t,x)\, O(x),
\end{align}
which means $V_I(t)=\int dx\, w(t,x)\, O^I(t,x)$.
We hereafter omit the label $I$ representing the interaction picture.
Let $O$ be a primary operator with a conformal dimension $\Delta$.
We note that similar computations are done and also relations to holography are investigated in literature (see, e.g., \cite{Berenstein:2014cia} and references therein).

From the results in the previous subsection, the leading changes of the expectation value of $O$ and $H_0$ are given by\footnote{The leading term \eqref{leadingO0} for $O$ vanishes in this case because the cylinder one-point function vanishes due to the conformal symmetry.}
\begin{align}
\label{CFTdeltaO}
    \delta \braket{O}(t,x)&=\int d^2x'\,   w(t',x') \theta (t-t')G(t-t',x-x'),
    \\
\label{CFTdeltaE}    
    \delta E(t)&=-\int d^2x' \,w(t',x')\theta (t-t')\frac{d\delta \braket{O}}{dt}(t',x')
\end{align}
with
\begin{align}
    G(t,x):=-i \eqb{[ O(t,x),O(0)]},
    \label{GCFT}
\end{align}
where we have assumed the translation invariance of time and space in $G$. As in the previous subsection, we have extended the integration region of $t'$ to $-\infty$ supposing that the source $w(t,x)$ vanishes before the initial time $t_i$ ($w(t,x)=0$ for $t<t_i$).

We consider the infinite line $-\infty<x<\infty$ where our CFT lives. 
The form of the two-point function is independent of the details of CFTs.
On the Euclidean cylinder with periodicity $\beta$, we have
\begin{align}\label{ECFT2pt}
     &\eqb{O(z_1,\bar{z}_1) O(z_2,\bar{z}_2)}
     = \left(\frac{\pi}{\beta}\right)^{2\Delta}
    \left(\sinh \left(\frac{\pi}{\beta}z_{12}\right)\sinh \left(\frac{\pi}{\beta}\bz_{12}\right)\right)^{-\Delta},\\
    &z_{12}=z_1-z_2, \qquad \bz_{12}=\bz_1-\bz_2\nonumber,
\end{align}
where $z, \bz$ is the complex coordinates of the cylinder $z=x+i\tau$ where $\tau$ is Euclidean time with periodicity $\beta$.  
Lorentzian time correlators can be obtained from the analytic continuation of the Euclidean correlator (see, e.g., \cite{Hartman:2015lfa, simmons2019tasi} for the $i\epsilon$-prescription in Lorentzian correlators). 
We obtain $G(t,x)$ defined in \eqref{GCFT} as
\begin{align}
   G(t,x)
   =-i \left(\frac{\pi}{\beta}\right)^{2\Delta}
   &\left[
   \left(\sinh \left(\frac{\pi (-u+i\epsilon)}{\beta}\right)\sinh \left(\frac{\pi (v-i\epsilon)}{\beta}\right)\right)^{-\Delta}
    \right.\nn
    &\qquad\left.
    - \left(\sinh \left(\frac{\pi (u+i\epsilon)}{\beta}\right)\sinh \left(\frac{\pi (-v-i\epsilon)}{\beta}\right)\right)^{-\Delta}
   \right],
   \label{Guv}
\end{align}
 where $u=t-x$, $v=t+x$.

Notice that $G(t,x)$ is singular at $u=0$ or $v=0$ in the limit $\epsilon \to 0$.
Due to this singular behavior, the response of the one-point function $\delta \braket{O}(t,x)$ given by \eqref{CFTdeltaO} diverges for $\Delta \geq 1$ if $w(t',x')$ has a support on the past light cone of $(t,x)$. 
For $\delta \braket{O}(t,x)$, we can avoid the divergence by considering only specific $(t,x)$ for given source $w(t',x')$. 
For example, if $w(t',x')$ takes a non-vanishing value only on a finite spacetime region $M$, then $\delta \braket{O}(t,x)$ does not diverge if $(t,x)$ is outside of the chronological future of $M$. 
However, the change of the energy $\delta E(t)$ diverges at every time $t>t_i$ for $\Delta \geq 1$. 
This is clear from \eqref{CFTdeltaE} where $d\delta \braket{O}/dt$ is singular at the point where $w(t',x')$ is non-zero.
We revisit this point in appendix \ref{app: retarded}.

The divergence of $\delta E$ is essential because the perturbation by the local operator \eqref{CFT:V(t)} contains arbitrarily high energy modes.
One way to regularize the divergence is by introducing the Euclidean-time damping factors as
\begin{align}
\label{CFT:Vreg}
    V_\text{reg}(t)=\int dx\, w(t,x)\, e^{-a H_0}O(x)e^{-a H_0}, \qquad (a>0)
\end{align}
where we added $e^{-a H_0}$ symmetrically so that $ V_\text{reg}(t)$ is Hermitian.
Then, we can make $\delta E(t)$ finite, by keeping $a$ non-vanishing.
For example, when $w(t,x)=\lambda \theta(t-t_i)\delta(x-x_0)$, we have
\begin{align}
    \delta E=\frac{Z(\beta')}{Z(\beta)}\lambda^2 2^{\Delta+1} \Delta \left(\frac{\pi}{\beta'}\right)^{2\Delta+1}
     \sin\left(\frac{4\pi}{\beta'}a\right)
    \left(1-\cos \left(\frac{4\pi}{\beta'}a\right)\right)^{-\Delta-1},
\end{align}
where $\beta'=\beta+4a$.
This is always positive because $0<4\pi a/\beta'=4\pi a/(\beta+4a)<\pi$ if $\beta>0$ and $a>0$.
Therefore, through \eqref{S=betaE}, the second law $\delta S \geq 0$ is satisfied. 

However, we have not carried out the bulk computations with the regularization corresponding to the Euclidean-time damping. 
Thus, in the following computation, we will just keep $\epsilon$ in \eqref{Guv}, and will see how $\delta E$ diverges in the limit $\epsilon \to 0$.

\subsubsection*{Spatially uniform source}
We now consider a spatially uniform source $w(t)$ as in the example in section~\ref{sec: example}.
In this case, $\delta \braket{O}$ given by \eqref{CFTdeltaO} is independent of $x$, and thus $\delta E$ given by \eqref{CFTdeltaE} is proportional to the volume of the space. 
To avoid this trivial IR divergence, we consider, instead of $\delta E$,  the change of the energy density $\delta \mathcal{E}$ which is given by
\begin{align}
    \delta \mathcal{E}(t)&:=-\int^{t}_{-\infty}\!\! dt'\,w(t')\frac{d\delta \braket{O}}{dt}(t').
    \label{def:del-calE}
\end{align}
Introducing 
\begin{align}\label{eq: uniform G}
    G(t) := \int\d x\, G(t,x),
\end{align}
where $G(t,x)$ is given by \eqref{Guv},
$\delta \mathcal{E}$ is written as (see \eqref{delEeq})
\begin{align}
    \delta \mathcal{E}(t)=-\int^{t}_{t_i}\!\!dt'\,\int^{t'}_{t_i}\!\!dt''\, w(t')w(t'')\frac{d}{dt'}G(t'-t'').
    \label{delE:G(t-t')}
\end{align}
Using this expression, we can numerically compute $\delta \mathcal{E}(t)$ for several source profiles $w(t)$.
In particular, if $w(t)$ is a rectangular function on the interval $[0,1]$ as
\begin{align}\label{eq: w rect}
    w_\text{rect}(t)=\begin{cases}
        1 & (0\leq t \leq 1)\\
        0 & (\text{otherwise})
     \end{cases},
\end{align}
we have
\begin{align}
     \delta \mathcal{E}(t)=-\int^{t}_{0}\!\!dt'\,G(t')
\end{align}
for $0<t<1$.
Here, we have set $t_i=0$.

\begin{figure}[p]
    \centering
    \begin{minipage}[b]{.45\textwidth}
        \centering
        \includegraphics[width=\textwidth]{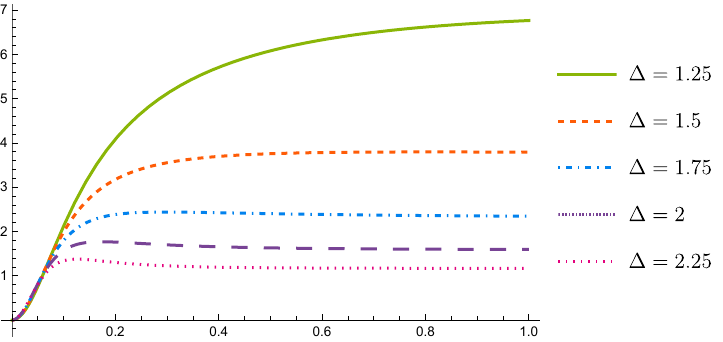}
    \end{minipage}%
    \hspace{0.04\textwidth}
    \begin{minipage}[b]{.45\textwidth}
        \centering
        \includegraphics[width=\textwidth]{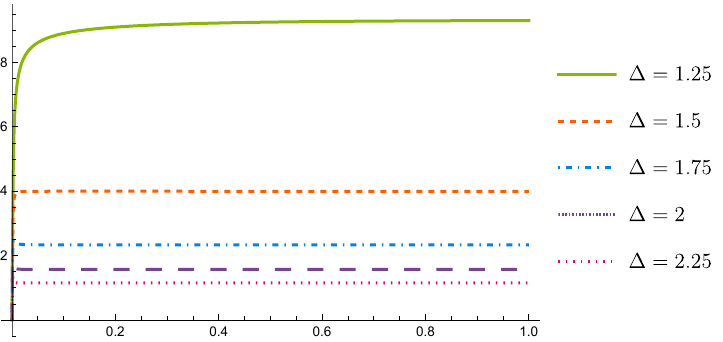}
    \end{minipage}
    \caption{Time-dependence of $\delta \mathcal{E}$ for $\Delta>1$ with $\epsilon=10^{-1}$ (left) and $\epsilon=10^{-3}$ (right). 
    The source $w(t)$ is given as \eqref{eq: w rect}.
    The vertical axis represents $\epsilon^{2\Delta-2}\delta\mathcal{E}(t)$, and the horizontal axis is $t$ with $\beta=\pi$.}
    \label{fig:delta>1}
\end{figure}

\begin{figure}[p]
    \centering
    \begin{minipage}[b]{.5\textwidth}
        \centering
        \includegraphics[width=\textwidth]{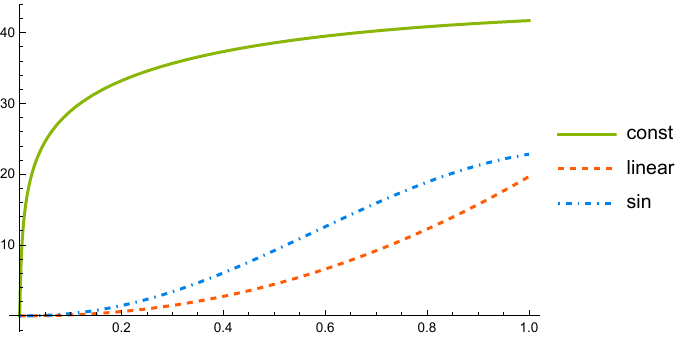}
    \end{minipage}%
    \hspace{0.04\textwidth}
    \begin{minipage}[b]{.35\textwidth}
        \centering
        \includegraphics[width=\textwidth]{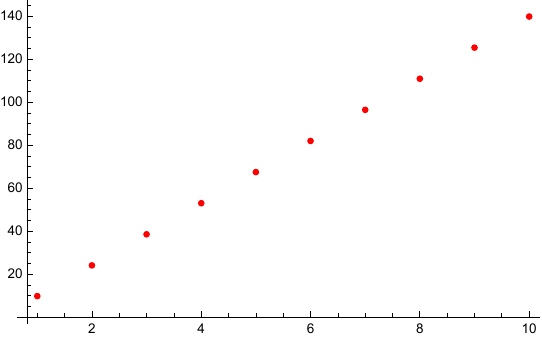}
    \end{minipage}
    \caption{(Left) Time-dependence of $\delta \mathcal{E}$ for $\Delta=1$ with various sources $w(t)=1$, $w(t)=t$, $w(t)=\sin(\pi t/2)$ in the interval $0\leq t\leq 1$. 
    The parameters are set as $\epsilon=10^{-3}$ and $\beta=\pi$.
    (Right) $\epsilon$-dependence of $\delta \mathcal{E}$ for $\Delta=1$. The vertical axis represents $\delta\mathcal{E}(1/2)$ with the constant source $w(t)=1$ for various $\epsilon=10^{-n}$ $(n=1,\dots, 10)$, and the horizontal axis is this $n$. This plot indicates that $\delta \mathcal{E}$ for $\Delta=1$ diverges as $\log \epsilon^{-1}$.}
    \label{fig:delta=1}
\end{figure}

\begin{figure}[p]
    \centering
    \begin{minipage}[b]{.5\textwidth}
        \centering
        \includegraphics[width=\textwidth]{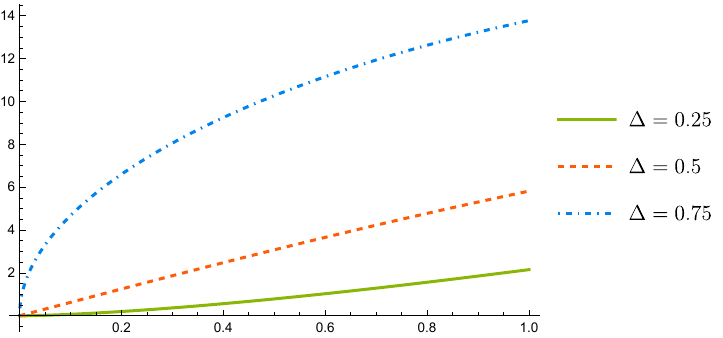}
    \end{minipage}%
    \hspace{0.04\textwidth}
    \begin{minipage}[b]{.35\textwidth}
        \centering
        \includegraphics[width=\textwidth]{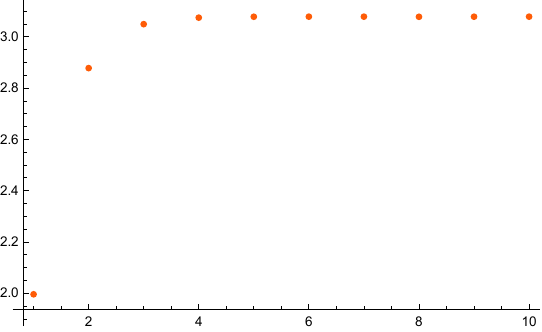}
    \end{minipage}
    \caption{(Left) Time-dependence of $\delta \mathcal{E}$ for $\Delta<1$ with $\epsilon=10^{-3}$. The source takes the constant value $w(t)=1$ (only) in the interval $0\leq t\leq 1$. The vertical axis represents  $\delta\mathcal{E}(t)$, and the horizontal axis is $t$ with $\beta=\pi$. Note that $\delta\mathcal{E}(t)$ is not rescaled by $\epsilon$ unlike the plots for $\Delta>1$ (Fig.\ref{fig:delta>1}). 
    (Right) $\epsilon$-dependence of $\delta \mathcal{E}$ for $\Delta=1/2$. The vertical axis represents $\delta\mathcal{E}(t)$ at $t=1/2$ with the constant source $w(t)=1$ for various $\epsilon=10^{-n}$ $(n=1,\dots, 10)$, and the horizontal axis is this $n$. This plot indicates that $\delta \mathcal{E}$ for $\Delta<1$ takes a finite value in the limit $\epsilon \to 0$.}
    \label{fig:delta<1}
\end{figure}

We give the numerical results of $\delta \mathcal{E}(t)$ (Fig.~\ref{fig:delta>1}, \ref{fig:delta=1}, \ref{fig:delta<1}). 
As we have discussed, $\delta \mathcal{E}$ may have UV divergences if we take the limit removing $\epsilon$ in \eqref{Guv}.
Thus, we keep $\epsilon$ finite in the numerical computations. 
The plots show that $\delta \mathcal{E}(t)$ is always non-negative if we keep $\epsilon$ finite.
As shown in the figures, the second law is satisfied.
In particular, for $\Delta<1$, $\delta \mathcal{E}$ is UV finite.

Fig.~\ref{fig:delta>1} shows the time-dependence of $\delta \mathcal{E}$ with $\Delta >1$ for the rectangular source $w_\text{rect}(t)$. 
These plots indicate that the leading UV divergence is proportional to $\epsilon^{2-2\Delta}$ for $\Delta>1$. 
Fig.~\ref{fig:delta=1} shows plots for $\Delta=1$. The right plot in Fig.~\ref{fig:delta=1} indicates that the UV divergence is proportional to $\log \epsilon^{-1}$ for $\Delta=1$.
Fig.~\ref{fig:delta<1} shows the plots for $\Delta<1$. The right plot in Fig.~\ref{fig:delta<1} represents that there are no UV divergences for $\Delta<1$, and we can take the limit $\epsilon \to 0$ safely. 

We now compare the CFT computations to the bulk ones.
Introducing the Fourier transform of $G$ as\footnote{\eqref{tildeG} is well-defined by keeping $\epsilon$. However, the inverse Fourier transformation is ill-defined. Owing to this fact, \eqref{delO:Fourier} is just a formal expression for $\Delta \geq 1$ unless we introduce some regularization in the $\omega$-integral.
We will give a more precise discussion in appendix~\ref{app: retarded}.}
\begin{align}
\label{tildeG}
    \tilde{G}(\omega)&:=\int d^2x \,e^{i\omega t} G(t,x),
\end{align}
we obtain
\begin{align}
    \delta \braket{O}(t)&=\int^{t}_{-\infty}\!\! d t'\,   w(t') \int^{\infty}_{-\infty}\!\! dx' G(t-t',x-x')
    \nn
    &=\int^{t}_{-\infty}\!\! d t'\,   w(t') \int \frac{d\omega}{2\pi}e^{-i\omega (t-t')}\tilde{G}(\omega).
    \label{delO:Fourier}
\end{align}
Using the explicit form of $G$ in \eqref{Guv}, we can compute $\tilde{G}(\omega)$ as
\begin{align}
       \tilde{G}(\omega)
    &=-\frac{i}{2}\left(\frac{\pi}{\beta}\right)^{2\Delta-2}
    \left[\left|I_-\left(\frac{\beta\omega}{2\pi}\right)\right|^2
    -\left|I_+\left(\frac{\beta\omega}{2\pi}\right)\right|^2\right],
\end{align}
where $I_{\pm}(y)$ is the following function:
\begin{align}
    I_\pm(y):=\lim_{\epsilon\to 0+}\int dx \frac{e^{iyx}}{(\sinh (x\pm i \epsilon))^{\Delta}}
    =2^{\Delta-1}e^{\mp\frac{\pi y}{2}\mp\frac{i\pi \Delta}{2}}\frac{\Gamma\left(\frac{\Delta+ i y}{2}\right)\Gamma\left(\frac{\Delta- i y}{2}\right)}{\Gamma(\Delta)}.
\end{align}
One can also find that $I_\pm(y)$ is written as
\begin{align}
   I_\pm(y)&= 2^{\Delta-1}\Gamma(1-\Delta)\left[A(y)+e^{\mp i \pi \Delta}A(-y)\right],
   \\
   A(y)&:=\frac{\Gamma\left(\frac{\Delta- i y}{2}\right)}{\Gamma\left(1-\frac{\Delta+ i y}{2}\right)}.
\end{align}
It leads to 
\begin{align}\label{eq: Fourier of Pauli-Jordan}
       \tilde{G}(\omega)
    &=-\left(\frac{2\pi}{\beta}\right)^{2\Delta-2}\sin\left(\pi\Delta\right)\Gamma(1-\Delta)^2
\left[A\left(\frac{\beta\omega}{2\pi}\right)^2-A\left(-\frac{\beta\omega}{2\pi}\right)^2\right]
\nn
&=-\left(\frac{2\pi}{\beta}\right)^{2\Delta-2}\frac{\pi\Gamma(1-\Delta)}{\Gamma(\Delta)}
\left[A\left(\frac{\beta\omega}{2\pi}\right)^2-A\left(-\frac{\beta\omega}{2\pi}\right)^2\right].
\end{align}

We then have
\begin{align}\label{eq: tilde G inverse FT}
    &\int \frac{d\omega}{2\pi}e^{-i\omega (t-t')}\tilde{G}(\omega)
    =-\left(\frac{2\pi}{\beta}\right)^{2\Delta-2}\frac{\pi\Gamma(1-\Delta)}{\Gamma(\Delta)}
    \int \frac{d\omega}{2\pi}e^{-i\omega (t-t')}A\left(\frac{\beta\omega}{2\pi}\right)^2.
\end{align}
Here, we have dropped $A(\beta\omega/(2\pi))^2$ term, because 
it does not contribute to this integral.
In fact, since $t-t'>0$ and $A(-\beta\omega/(2\pi))^2$ is analytic in the lower half plane for $\Delta>0$, we can change the integration contour to the complex lower half plane.
However, we must note that for $\Delta \ge 1$ the $\omega$-integral is ill-defined originally, as the integrand diverges at the infinity, which is confirmed by Stirling's formula.
We will come back to this point and discuss to what extent our analysis can be validated in appendix~\ref{app: retarded}.
Keeping it in mind, we further proceed to the linear response:
\begin{align}
    \delta \braket{O}(t)
    &=-\left(\frac{2\pi}{\beta}\right)^{2\Delta-2}\frac{\pi\Gamma(1-\Delta)}{\Gamma(\Delta)}
    \int \frac{d\omega}{2\pi} A\left(\frac{\beta\omega}{2\pi}\right)^2\int^{t}_{-\infty}\!\! d t'\,  e^{-i\omega (t-t')} w(t')
    \nn
    &=-\left(\frac{2\pi}{\beta}\right)^{2\Delta-2}\frac{\pi\Gamma(1-\Delta)}{\Gamma(\Delta)}
    \int \frac{d\omega}{2\pi} A\left(\frac{\beta\omega}{2\pi}\right)^2\int^{\infty}_{-\infty}\!\! d t'\,  e^{-i\omega (t-t')} w(t'),
    \label{delO_CFT_omega}
\end{align}
where we have extended the integration region of $t'$ because the $\omega$-integral vanishes if $t-t'<0$ since $A(\beta\omega/(2\pi))^2$ is analytic in the upper half $\omega$-plane.

The result \eqref{delO_CFT_omega} agrees with \eqref{eq: final Pi} in the bulk computations, where to see this we use $r_+ = 2\pi L^2/\beta$.
The difference of the positive dimensionless overall factors is just a matter of the normalization of $O$ and can be absorbed into $C$ in \eqref{eq: final Pi}.
In the bulk computations, the space where the dual CFT lives can be a compactified space, while we have considered a CFT on the infinite line. 
The reason why the two results are the same is because the source is spatially uniform. 
Indeed, for holographic CFT on a circle with circumference $\ell$, the two-point function $G_\ell(t,x)$ can be computed from $G(t,x)$ on the infinite line \eqref{Guv} by the method of images (see e.g. \cite{Keski-Vakkuri:1998gmz, Maldacena:2001kr, Birmingham:2002ph}) as
\begin{align}
    G_\ell(t,x)=\sum_{n=-\infty}^\infty G(t,x+n\ell).
\end{align}
Then, if we smear it with the uniform source on the circle, it agrees with the smeared one on the infinite line as
\begin{align}
    \int^\ell_0 \!\! dx\,G_\ell(t,x)=\int^\infty_{-\infty}\!\! dx\,G(t,x).
\end{align}
If we consider a space-dependent source on a compactified space, the result depends on the theory because torus two-point functions depend on the details of CFTs.

From \eqref{delO_CFT_omega}, the change of the energy density $\delta \mathcal{E}$ defined in \eqref{def:del-calE} can be represented as 
\begin{align}
    \delta \mathcal{E}(t)=-i\left(\frac{2\pi}{\beta}\right)^{2\Delta-2}\frac{\pi\Gamma(1-\Delta)}{\Gamma(\Delta)}
    \int \frac{d\omega}{2\pi} \omega A\left(\frac{\beta\omega}{2\pi}\right)^2\int^{t}_{-\infty}\!\! dt'\,w(t')\int^{\infty}_{-\infty}\!\! d t''\,  e^{-i\omega (t'-t'')} w(t'').
\end{align}

\section{Retarded Green's function as a distribution}\label{app: retarded}
Here, we give a proof that the retarded Green's function obtained in the CFT and that in the bulk are equivalent only when the Green's function is seen as a distribution whose class of test functions are chosen properly.
By this, \eqref{eq: final Pi} is guaranteed to be equal to \eqref{eq: CFT one point} after a time that the source $w(t)$ is off.

We start with the CFT side and define the retarded Green's function for the $s$-wave as
\begin{align}\label{eq: G_R before dist}
    G_R(t) := \int\d x\, G_R(t,x)= \Theta(t) \int\d x\, G(t,x),
\end{align}
where $G(t,x)$ is introduced in \eqref{Guv}.
The limit of $\epsilon \to 0$ is always understood to be taken after all the calculation.
In this sense, $G_R$ can be regarded as a distribution.
Since it diverges near $t\approx 0$ (the contact point of the correlator), we consider the class of the test functions which have support on $t>0$ and decay at $t=0$ as rapidly as $e^{-1/t}$.
Furthermore, they are assumed to be Schwartz functions to ensure that the Fourier transform of $G_R$ and its inverse are well-defined.
In this setup, $G_R$ is defined as a map that maps a test function $\varphi$ to
\begin{align}\label{eq: G_R as dist}
    G_R[\varphi] := \lim_{\epsilon\to 0+} \int_{-\infty}^\infty \d t\, G_R(t)\varphi(t).
\end{align}

The Fourier transform $\tilde G_R$ is mathematically defined as a distribution such that
\begin{align}\label{eq: tilde G_R as dist}
    \tilde G_R[\tilde \varphi] = G_R[\varphi],\qquad
    \mbox{with}\qquad
    \tilde \varphi(\omega) := \int_{-\infty}^\infty\frac{\d t}{2\pi}e^{-i\omega t}\varphi(t)
    =
    \int_{0}^\infty\frac{\d t}{2\pi}e^{-i\omega t}\varphi(t).
\end{align}
Following this definition, we find $\tilde G_R$ as follows.
First, \eqref{eq: G_R as dist} is equivalent to
\begin{align}
    G_R[\varphi] = \lim_{\epsilon\to 0+} \int_{-\infty}^\infty \d t\, G(t)\varphi(t),
\end{align}
where $G$ is defined in \eqref{eq: uniform G} and we have used the fact that the support of $\varphi$ is $(0,\infty)$.
Then, using \eqref{eq: Fourier of Pauli-Jordan}, we obtain
\begin{align}
    G_R[\varphi] &= \lim_{\epsilon\to 0+} \int_{-\infty}^\infty \d t\, G(t)\int_{-\infty}^\infty\d \omega\,e^{i\omega t}\tilde{\varphi}(\omega)\nonumber\\
    &= \int_{-\infty}^\infty\d \omega\, \tilde{\varphi}(\omega)\lim_{\epsilon\to 0+}\int_{-\infty}^\infty \d t\, e^{i\omega t}G(t)
    = \int_{-\infty}^\infty\d\omega\,\tilde G(\omega) \tilde\varphi(\omega).
\end{align}
Since $\tilde\varphi(\omega)$ vanishes rapidly at the lower infinity by definition \eqref{eq: tilde G_R as dist}, the term of $A(-\beta \omega/(2\pi))^2$ in \eqref{eq: Fourier of Pauli-Jordan}, which is analytic in the lower half plane, does not contribute in the above last integral.
This leads to
\begin{align}
    G_R[\varphi] &=-\left(\frac{2\pi}{\beta}\right)^{2\Delta-2}\sin\left(\pi\Delta\right)\Gamma(1-\Delta)^2\int_{-\infty}^\infty\d \omega A\left(\frac{\beta\omega}{2\pi}\right)^2\tilde\varphi(\omega).
\end{align}
Now we can see this as a map for $\tilde\varphi$, and hence it is identified as $\tilde G_R[\tilde\varphi]$.
Note that the last expression would not be convergent if $\varphi$ were an arbitrary function.
Schematically written, the Fourier transform is
\begin{align}
    \tilde G_R(\omega) = -\left(\frac{2\pi}{\beta}\right)^{2\Delta-2}\sin\left(\pi\Delta\right)\Gamma(1-\Delta)^2 A\left(\frac{\beta\omega}{2\pi}\right)^2,
\end{align}
which coincides with the bulk result \eqref{eq: final Pi} with $w_\omega \to 1$ and choosing the normalization  factor $C$ appropriately.

So, what can we conclude about the one-point function $\braket{O}_t$?
In \eqref{eq: CFT one point}, $G_R(t)$ appears as $G_R(t-t')$, and thus $w(t')$ must go to zero at $t'=t$ to make $\braket{O}_t$ finite, meaning that $\braket{O}_t$ is well-defined only after the source is turned off.
Then, under the promise that the support of $w$ is compact, the Fourier transform is defined as we did above after $w$ is turned off, and we can conclude $\braket{O}_t = \Pi(t)$ for any $\Delta>0$ with the constants identified properly between the boundary and bulk.

Unfortunately, however, the energy and entropy at time $t$ require all the history until $t$ as \eqref{delEeq}, making it inevitable to use the value of the one-point function at $t$ when $w$ is still non-zero.
In other words, the integration in \eqref{delE:G(t-t')} always contains the coincident points $t'=t''$ for any choice of the source $w(t)$.
Then, the equivalence between $\Pi(t)$ and $\braket{O}_t$ is no longer guaranteed, and as a result, we reached the conclusion that the second law is broken when we use $\Pi(t)$ in \eqref{eq: final Pi} for $\Delta>1$.
The reason why the second law is satisfied for $0<\Delta<1$ is now easily understood; $G_R(t)$ in \eqref{eq: G_R before dist} is finite everywhere in this case, so the Fourier transform is well-defined without carefully specifying the class of the test functions.
For $\Delta>1$, we need to discover a way to recover the retarded Green's function with a regulator also in the bulk, or a renormalization scheme compatible with the second law.
The latter may be better because the entropy becomes independent of the UV regulator.
We hope that our second law serves as a guiding principle in identifying a good renormalization that respects the non-negativity of relative entropy, one of the most fundamental properties of quantum mechanics.

\bibliographystyle{jhep} 
\bibliography{ref}
\end{document}